\documentclass[fleqn,usenatbib]{mnras}

\usepackage{newtxtext,newtxmath}
\usepackage{multicol}

\usepackage[T1]{fontenc}
\usepackage{ae,aecompl}

\usepackage{graphicx}	
\usepackage{amsmath}	
\usepackage{amssymb}	
\usepackage{threeparttable}
\usepackage{float}

\title[Photospheric and chromospheric activity on SV Cam]{Star-spot distributions and chromospheric activity on the RS CVn type eclipsing binary SV Cam\thanks{This paper includes data taken at the Calar Alto, the Roque de los Muchachos observatories and the Astronomical Observatory of the Jagiellonian University.}}

\author[H.V. \c{S}enavc{\i}]{
H.V. \c{S}enavc{\i},$^{1}$\thanks{E-mail: hvsenavci@ankara.edu.tr}
E. Bahar,$^{1}$
D. Montes, $^{2}$
S. Zola, $^{3}$
G.A.J. Hussain, $^{4}$
A. Frasca, $^{5}$
\newauthor{
E. I\c{s}{\i}k,$^{6,7}$
and {O}. Y\"{o}r\"{u}ko\u{g}lu $^{1}$}
\\
$^{1}$Ankara University, Faculty of Science, Department of Astronomy and Space Sciences, Tandogan, Ankara, Turkey\\
$^{2}$Dpto. Astrofisica, Facultad de CC. Fisicas, Universidad Complutense de Madrid, Spain\\
$^{3}$Astronomical Observatory, Jagiellonian University, ul. Orla 171, PL-30-244 Krakow, Poland\\
$^{4}$European Southern Observatory, Karl-Schwarzschild-Str. 2, 85748 Garching bei M\"unchen, Germany\\
$^{5}$INAF -- Osservatorio Astrofisico di Catania, via S. Sofia, 78, 95123 Catania, Italy\\
$^{6}$Max-Planck-Institut f\"{u}r Sonnensystemforschung, Justus-von-Liebig-Weg 3, 37077, G\"{o}ttingen, Germany\\
$^{7}$Feza G\"ursey Center for Physics and Mathematics, Bo\u{g}azi\c{c}i University, Kuleli/\"Usk\"udar, 34684, Istanbul, Turkey \\
}

\date{Accepted XXX. Received YYY; in original form ZZZ}

\pubyear{2015}
\begin{document}
\label{firstpage}
\pagerange{\pageref{firstpage}--\pageref{lastpage}}
\maketitle

\begin{abstract}Using a time series of high-resolution spectra and high-quality multi-colour photometry, we reconstruct surface maps of the primary component of the RS CVn type rapidly rotating eclipsing binary, SV Cam (F9V + K4V). We measure a mass ratio, $q$, of $0.641(2)$ using our highest quality spectra and obtain surface brightness maps of the primary component, which exhibit predominantly high-latitude spots located between 60$^{\circ}$ - 70$^{\circ}$ latitudes with a mean filling factor of $\sim$35\%. This is also indicated by the R-band light curve inversion, subjected to rigourous numerical tests. The spectral subtraction of the H$_{\alpha}$ line reveals strong activity of the secondary component. The excess H$_{\alpha}$ absorption detected near the secondary minimum hints to the presence of cool material partially obscuring the primary star. The flux ratios of Ca II IRT excess emission indicate that the contribution of chromospheric plage regions associated with star-spots is dominant, even during the passage of the filament-like absorption feature.

\end{abstract}

\begin{keywords}
stars: activity -- (stars:) binaries: eclipsing -- stars: imaging
\end{keywords}



\section{Introduction}


RS CVn-type binaries are composed of F-K type dwarf/giant components. Their orbital periods most typically range from 1 day to 20 days. Systems with smaller orbital periods exhibit strong magnetic activity, which is thought to be related to rapid rotation. They offer laboratories to study stellar activity in post-main-sequence stars influenced by tidal effects \citep{Strassmeier2009}. The presence of cool spots on eclipsing RS CVn-type systems is responsible for significant variability in their light curves outside eclipses \citep{Berdyugina2005}. Their effects can also be detected in spectral line cores which form in the chromosphere, e.g., Ca~II H\&K  and H$_\alpha$, where core emission can be much stronger than in the Sun \citep{Strassmeier2000}. 

SV Cam is a detached close binary of RS CVn-type (HD~44982, $m_{v}$=8.40), with an orbital period of $\sim$ 0.593 days. There have been several estimates of its spectral type, ranging from F9V to G2V for the primary component and K2V to K7V for the secondary, respectively, the most recent one being F9V + K4V by \citet{Jeffers2006b}. The latest investigation on the orbital period variation of SV Cam were also performed by \citet{Manzoori2016}, indicating two cyclic variations that are interpreted as light-time effect (LITE) due to the existence of a third body and magnetic activity cycle of the system. There are also several studies in the literature related to the activity behavior of SV Cam. Based on photometric data spanning over 50 years, \citet{Zeilik1988} pointed out that the spot groups on the primary component fall into two zones as active longitude belts. Using the photometric observations  made by \citet{Patkos1982} during the 1973-1981 period, \citet{Djurasevic1998} performed spot modeling and mentioned the existence of active longitudes. In a series of papers, Zboril \& Djura{\v s}evi{\v c} (\citeyear{Zboril2003, Zboril2004, Zboril2006}) monitored the brightness variation of the system outside eclipses using data obtained between 2001 - 2005, and also argued about the existence of two spots on the primary component, but with varying dimensions and locations from one season to another. \citet{Hempelmann1997} obtained Doppler images of the system using Ca~I (6103, 6122 and 6439 $\mbox{\AA}$) and Fe I (6400 $\mbox{\AA}$) lines. The surface maps obtained in that study revealed a dominant spot region located at around 40 - 75$^{\circ}$ longitudes and latitudes between +30 and 60$^{\circ}$. \citet{Rucinski2002} determined the mass ratio of the system as q=$M_{2}$/$M_{1}$=0.641 and the spectral type of the primary component as G2 V. They also determined the masses of the primary and the secondary components as $1.14 M_{\odot}$ and $0.73 M_{\odot}$, respectively. With the help of spectroscopic observations covering the entire orbital period of SV Cam, \citet{Kjurkchieva2002} determined the mass ratio of the system as q=$M_{2}$/$M_{1}$=0.593. Their results indicated the presence of two cool spots on the primary component, with their maximum visibilities at phases 0.27 and 0.86. 

A series of papers concerning the star-spot distribution in SV Cam were published by \cite{Jeffers2005,Jeffers2006a,Jeffers2006b}. In \citet{Jeffers2005}, they modeled the spectrophotometric HST data of SV Cam using PHOENIX model atmospheres and found that the surface flux in the eclipsed low-latitude region of the primary is about 30\% lower than that computed from the models. They also investigated the full surface flux of the primary component concluding that there is an additional flux deficit from the entire primary, which can be explained via a large polar spot on this star, extending from the pole to the latitude of $\sim$ 48$^{\circ}$. Based on the spectrophotometric HST observations, they determined the effective temperatures of the primary and secondary components to be $T_{1}$ = 6013 K and $T_{2}$ = 4804 K, respectively. The study by \citet{Jeffers2006a} includes an eclipse-mapping of the primary component, aimed at determination of the filling factors and sizes distribution of star-spots, too small to be resolved by DI. In addition to the HST data they used ground-based photometric observations, in order to obtain the light curve of the system covering the full orbital cycle. Taking the advantage of high precision HST data, the detected strong discontinuities at the four contact points in the residuals of the fit to the light curve can only be removed by the reduction of the photospheric temperature and presence of a polar spot. They concluded that the spottedness of the stellar surface can have a significant impact on the determination of the stellar parameters, such as radius and effective temperature, in the components of a binary like SV Cam case. Using the first and second derivatives of the HST data, \citet{Jeffers2006b} determined the best-fitting atmosphere model by adjusting the models to the brightness variations during the primary eclipse. They also emphasized the importance of other stellar parameters such as limb darkening coefficients, as they can alter the intensity values across the stellar disc.

The most recent activity-related study of the system was performed by \citet{Manzoori2016}, who analyzed the  AAVSO (American Association of Variable Star Observers) 2006 - 2009 light curves of SV Cam. He mentioned that the star spots appear at high latitudes on both components of SV Cam. He obtained cyclic orbital period variations with periods of 23.3 and 20.2 years from the O-C analysis. Furthermore, the Fourier transform of the second quadrature brightness variations led to another possible periodicity of  $\sim$ 35 years.

In this work, we aim to study the activity behavior of SV Cam, combining several widely used tools together for the first time. We analyze the time-series high resolution spectroscopic data via Doppler imaging, spectral synthesis and subtraction techniques, supported by simultaneous light and radial velocity curve analysis to obtain accurate system parameters. Such well-focused, high-resolution studies of a given RS CVn star contributes not only as another snapshot in its long-term activity behavior, but also serves to improve our understanding of the atmospheric phenomenology in rapidly rotating cool stars.

\section{Observations and Data Reduction}

The high resolution time-series spectra of SV Cam were obtained between 29 and 31 October, 2015 with the CAFE spectrograph \citep{Aceituno2013} attached to the 2.2~m telescope at the Calar Alto Observatory (Almeria, Spain) and with the HERMES spectrograph \citep{Raskin2011} attached to the 1.2~m Mercator telescope at the Roque de los Muchachos Observatory (La Palma, Spain) between 12 and 18 January, 2017. At Calar Alto we acquired 7 CAFE spectra of the system that cover the wavelength range 4050\,\AA--9095\,\AA\  with an average resolution of R=62\,000. Each spectrum was acquired with an exposure time of 1500 seconds that gives signal-to-noise (SNR) values between $\sim$ 60 and 90. During the observing run in La Palma, we obtained 23 HERMES spectra of SV Cam with an average resolution of R=85\,000 and a wavelength coverage between 3780 and 9007 $\mbox{\AA}$. Seventeen of these spectra have SNR between $\sim$ 70 and 100. Six spectra were discarded as they have SNR $<$ 40 that makes them unsuitable for the subsequent analysis. The log of observations is given in Table~\ref{tab:table1}. The reduction of CAFE spectra, which includes bias subtraction and flat field correction, removal of cosmic rays, wavelength calibration and Heliocentric velocity correction was performed using the IRAF\footnote{IRAF is distributed by the 
National Optical Astronomy Observatory, which is operated by the Association of the Universities for Research in 
Astronomy, inc. (AURA) under cooperative agreement with the National Science Foundation.} (Image Reduction and Analysis Facility) standard
packages. The HERMES spectra were reduced with the automatic pipeline of the spectrograph \citep{Raskin2011}. Normalization of spectra was performed via a Python code developed by our working group. This code can perform normalizations using polynomial and spline fits with the help of the synthetic spectra generated from model atmospheres. During our observing runs, three slowly-rotating and non-active template stars, namely HD~143761 (G0~V), HD~4628 (K2.5~V) and HD~22049 (K2~V), were also observed.

\begin{figure}
	\includegraphics[width=20pc,height=20pc]{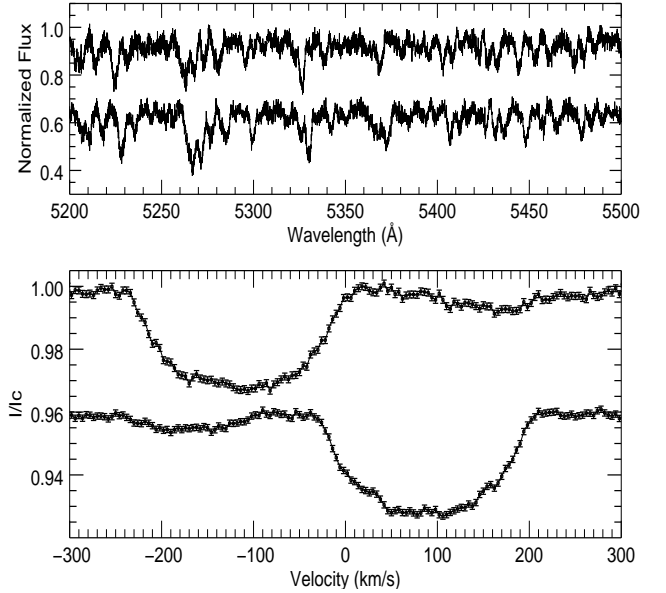}
 	\caption{The top panel shows two input spectra taken at orbital
 phases 0.32 and 0.65, while the lower panel shows the corresponding LSD profiles. Note that the strong and the weak dips in the LSD profiles belong to primary and secondary components, respectively.}
    \label{fig:figure1}
\end{figure}

\begin{figure}
\centering
	\includegraphics[width=15pc,height=20pc,angle=90]{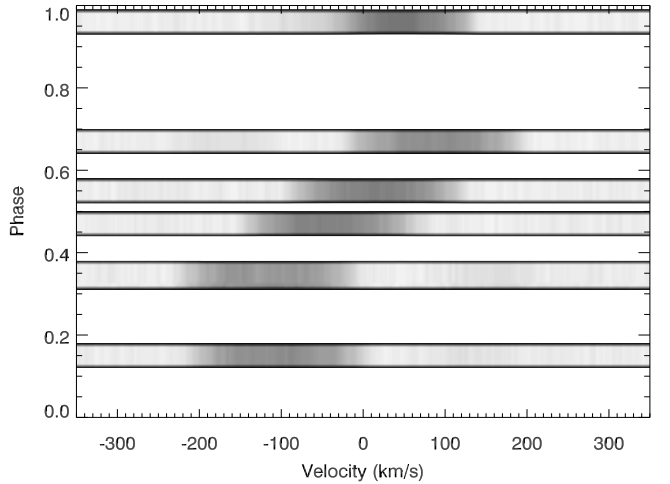}
 	\caption{A gray-scale image representation of Calar Alto dataset obtained using the LSD profiles of SV Cam.}
    \label{fig:figure2}
\end{figure}

\begin{figure}
\centering
	\includegraphics[width=15pc,height=20pc,angle=90]{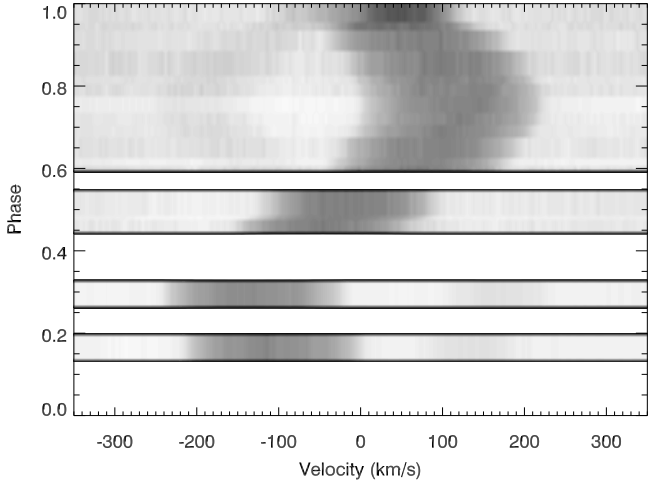}
 	\caption{Same as Figure~\ref{fig:figure2} but for La Palma dataset.}
    \label{fig:contrv_lap}
\end{figure}

In order to determine better the spot signatures in the stellar spectra, we enhanced the signal using the LSD technique \citep{Donati1997} and obtained mean velocity profiles with SNR values between 800 and 1100 for Calar Alto and La Palma datasets. The input lines list with line profile depths, that is required by the LSD technique, was extracted from the Vienna Atomic Line Database (VALD) \citep{Kupka1999}, by considering the $\log g$ and $T_{\rm eff}$, appropriate for SV Cam. During the preparation of the line list, wavelength regions covering lines affected by chromospheric heating (e.g. Ca II H\&K, H$\alpha$, Na D) and strong telluric lines were discarded in order to prevent artifacts in the LSD profiles. In addition to SV~Cam, the LSD profiles of the template stars were also computed and subsequently used for the generation of the lookup tables (used to model the local intensity profile). We show a sample of input spectra from Calar Alto observations and their corresponding LSD profiles in Figure~\ref{fig:figure1}. Time-series plots of the LSD profiles showing the data coverage as well as the radial velocity variations are shown in Figures~\ref{fig:figure2} and ~\ref{fig:contrv_lap} for the Calar Alto and La Palma datasets, respectively. The radial velocity variation of the secondary component is barely visible due to the weak contribution of that component to the total flux.

High precision $B$, $V$, $R$, $I$, and Str\"{o}mgren $b$ and $y$ photometry of SV Cam was obtained using the iKon L 936 Andor CCD camera attached to the 50-cm Cassegrain telescope at the Astronomical Observatory of the Jagiellonian University during two nights (31 December 2015 and 1 January 2016). We obtained 789, 721, 1488, 2116, 776 and 771 individual measurements in $B$, $V$, $R$, $I$, $b$ and $y$ bands, respectively. We chose GSC~4538-723 (K0~V)  as the comparison star to derive the differential magnitudes. We also observed TYC 4538-705-1 (B - V = 0.84) as check star, in order to check the variability of the comparison. The reduction procedure concerning the correction of scientific images for bias, dark and flat-field  was performed with IRAF while the C-Munipack software \citep{Hroch1998}, an interface for the DAOPHOT package, was applied for derivation of magnitudes.

\begin{table}
	\centering
	\caption{Phase ordered spectroscopic observation log of SV Cam.}
	\label{tab:table1}
	\begin{tabular}{lcccc} 
		\hline
	Date & Exp. Time  & HJD$_{Mid}$ & Phase$_{Mid}$ & SNR\\
    	&	(sec.)&	&	&\\
		\hline
		\multicolumn{5}{|l|}{Calar Alto Observing run}\\
		\hline
		29.10.2015 & 1500 & 57325.42262 & 0.143 & 61\\
		30.10.2015 & 1500 & 57325.53672 & 0.324 & 78\\
		31.10.2015 & 1500 & 57326.72813 & 0.333 & 64\\
		30.10.2015 & 1500 & 57325.61552 & 0.457 & 71\\
		30.10.2015 & 1500 & 57325.65956 & 0.531 & 79\\
		30.10.2015 & 1500 & 57325.73145 & 0.652 & 91\\
		31.10.2015 & 1500 & 57326.50529 & 0.945 & 84\\
		\hline
		\multicolumn{5}{|l|}{La Palma Observing run}\\
		\hline
		17.01.2017  &  1800  &  57771.41699  &  0.144 &  97\\
        17.01.2017  &  1800  &  57771.49726  &  0.280 &  104\\
        15.01.2017  &  1800  &  57768.63556  &  0.455 &  81\\
        18.01.2017  &  1800  &  57771.61709  &  0.482 &  96\\
        16.01.2017  &  1800  &  57770.43934  &  0.496 &  83\\
        14.01.2017  &  1500  &  57767.53940  &  0.606 &  69\\
        18.01.2017  &  1800  &  57771.69724  &  0.617 &  99\\
        14.01.2017  &  1500  &  57767.55778  &  0.637 &  74\\  
		17.01.2017  &  1800  &  57770.55080  &  0.684 &  91\\
        15.01.2017  &  1800  &  57769.38419  &  0.717 &  88\\
        18.01.2017  &  1800  &  57771.77563  &  0.749 &  102\\
        14.01.2017  &  1600  &  57767.64259  &  0.780 &  72\\
        14.01.2017  &  1600  &  57767.66169  &  0.813 &  74\\
        17.01.2017  &  1800  &  57770.64282  &  0.839 &  68\\
		15.01.2017  &  1800  &  57769.48775  &  0.892 &  71\\
        14.01.2017  &  1600  &  57767.73763  &  0.941 &  70\\
		17.01.2017  &  1800  &  57770.71976  &  0.969 &  71\\ 		  
		\hline
	\end{tabular}
\end{table}

\section{Orbital Solution and Photometric Analysis}

\begin{figure}
\centering
	\includegraphics[trim=1 1 1 1,clip,width=\columnwidth]{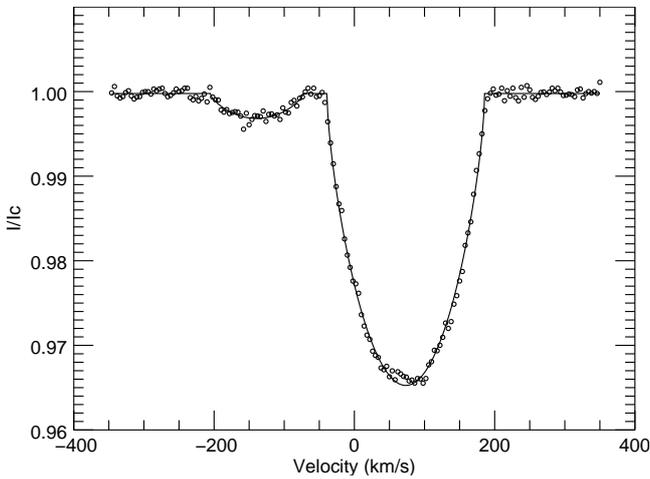}
 	\caption{An example of LSD profile (open circles) at phase = 0.60 obtained from La Palma observations and the synthetic LSD fit (solid line) used in radial velocity determination.}
    \label{fig:syntlsd}
\end{figure}

Eclipsing binary stars provide a unique opportunity to obtain the stellar masses directly as well as the orbital parameters, when both photometric and spectroscopic data are available. We first determined the radial velocities (RV) for each component of the system using both Calar Alto and La Palma spectra. As a first attempt, we used the IDL routines based on the broadening functions method improved by \citet{Rucinski1999}. However, since the flux contribution of the secondary component to the total flux of the system is very small, we could not estimate  precisely the RVs of the secondary component. The LSD technique, on the other hand, is a powerful tool which uses thousands of photospheric lines to enhance the SNR of the spectra. Consequently, we determined the radial velocities of both components of SV Cam by fitting synthetic rotation profiles to the LSD profiles with the help of the equation (equation 1) given by \citet{Barnes2004}. This equation is the improved version of the form (equation 17-12) given by \citet{Gray1992}, which is based on the determination of a rotationally broadened profile via the convolution of a non-rotating star profile and a rotation profile by considering the limb darkening effect. The modified version of this equation given by \citet{Barnes2004} additionally enables the adjustment of continuum level, width, depth and the radial velocities of each profile. An example of a fitted profile obtained in the radial velocity determination process is presented in Figure~\ref{fig:syntlsd}. We applied the related technique individually for both spectroscopic datasets. 

The simultaneous light curve and radial velocity analysis of SV Cam was performed with the PHOEBE \citep{Prsa2005} code.  We assumed a detached system configuration and carried out computations simultaneously for the RV data from La Palma and Calar Alto observations. We set the gravity darkening and the albedo coefficients to 0.32 \citep{Lucy1967} and 0.5 \citep{Rucinski1969}, respectively, which are theoretical values for a convective envelope appropriate for the spectral types of both components. The limb-darkening coefficients were interpolated automatically from the \citet{Hamme1993} tables under the linear-cosine law assumption. We fitted the orbital inclination (\textit{i}), effective temperature of the secondary component ($T{\rm _2}$), surface potentials of components (${\rm \Omega_{1,2}}$), the luminosity of the primary component ($L{\rm _1}$), the system mass ratio ($q=M{\rm _2}/M{\rm _1}$), semi-major axis ($a$) and the gamma velocity (V$_{\gamma}$). There are several studies in the literature concerning the effective temperature of the primary component of SV Cam system ranging from 5700 K \citep{Patkos1994} to 6440 K \citep{Albayrak2001}. \citet{Jeffers2006b} used the spectrophotometric HST data and determined the primary effective temperature as 6039 K $\pm$ 58 K. Consequently, we fixed the primary temperature to that value in our computations.

\begin{figure}
\centering
\includegraphics[trim=1 1 1 1,clip,width=\columnwidth]{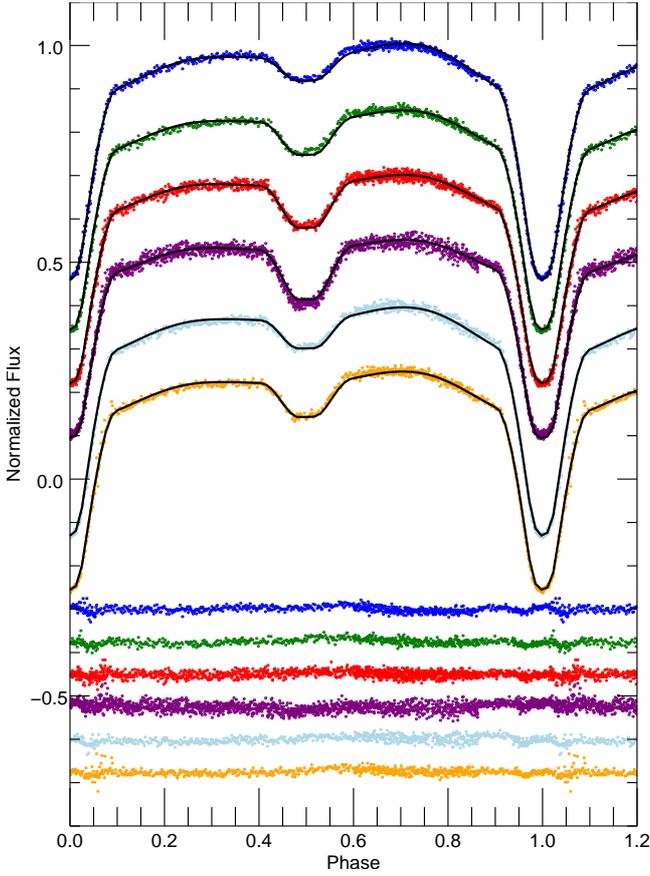}
\caption{$BVRIby$ light curves of SV Cam. Blue, green, red, purple, light blue and orange dots represent the observed light curves in the $B$, $V$, $R$, $I$, $b$ and $y$ filters, respectively, while the solid lines show best fits obtained from the simultaneous analysis using La Palma RV curves and the multi-band photometry. 
Except the ones related with B Band, all the observed and synthetic light curves are shifted arbitrarily for clarity by 0.15, in relative flux units, and the residuals by 0.075.}
    \label{fig:figure5}
\end{figure}

\begin{figure}
\centering
	\includegraphics[trim=1 10 1 60,clip,width=\columnwidth]{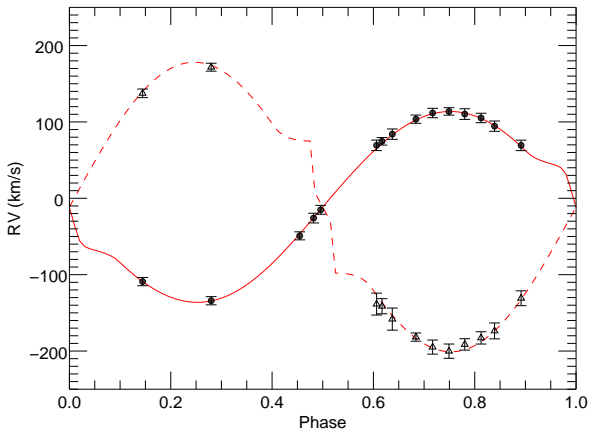}
 	\caption{Radial velocity curve of SV Cam obtained using HERMES spectra. The solid and dashed red lines represent the RV fit to the data, while the filled circles and open triangles belong to the radial velocity data of the primary and secondary components, respectively.}
    \label{fig:rvfitlp}
\end{figure}

\begin{figure}
\centering
	\includegraphics[trim=1 10 1 30,clip,width=\columnwidth]{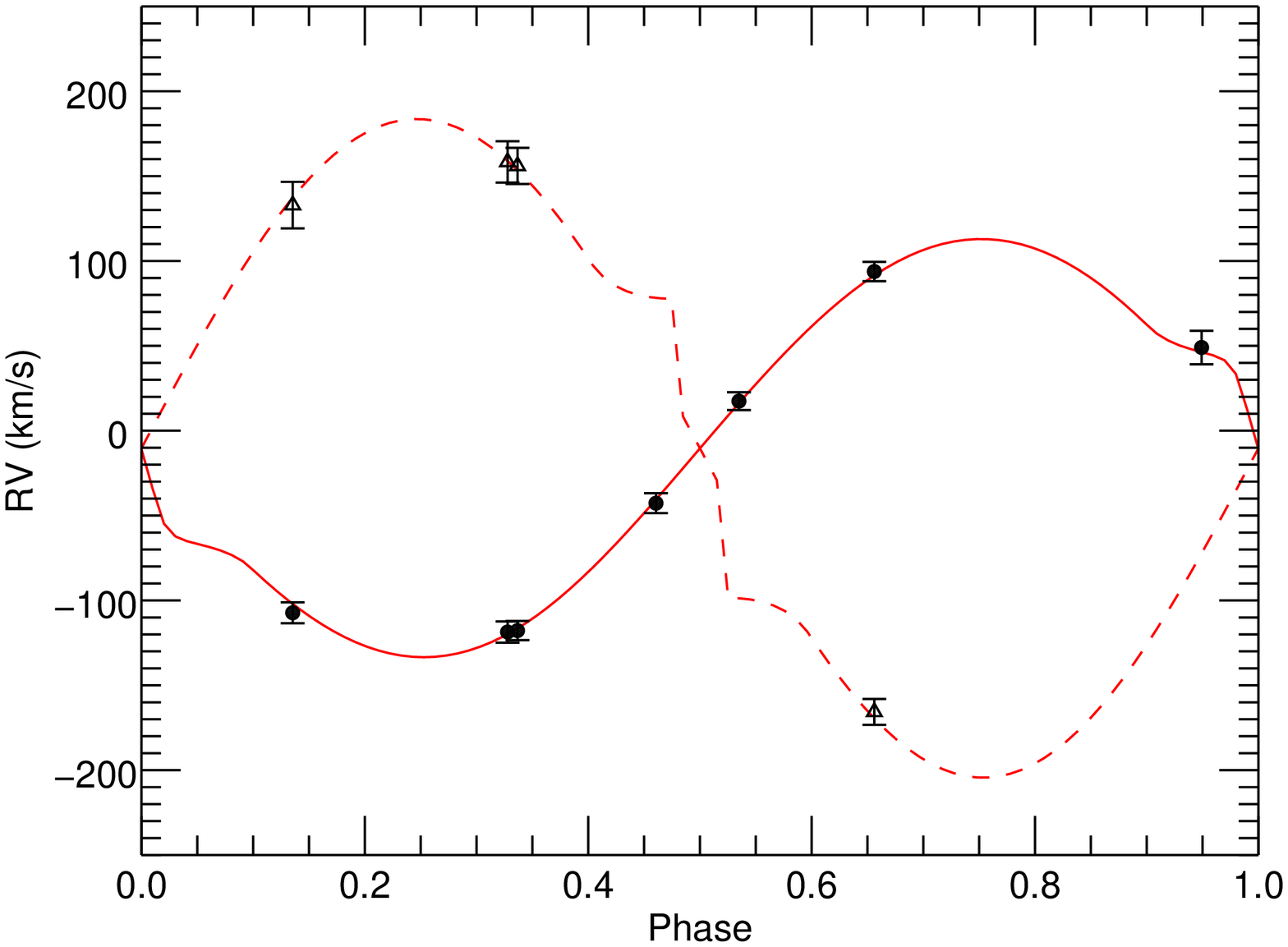}
 	\caption{Same as Figure~\ref{fig:rvfitlp} but for the CAFE (Calar Alto) dataset.}
    \label{fig:rvfitca}
\end{figure}

The light and radial velocity curve analysis of SV Cam showed that convincing solutions can only be achieved via cool spot modeling. Most of the photometry-based spot modelling studies published in the literature mentioned the presence of one or two cool spots with their locations and sizes being variable from one season to another. Therefore, we performed the spot modeling by assuming one, two and even three cool spots located on the surface of the primary component and adjust all the spot-related parameters (co-latitude, longitude, radius and temperature factor) during the iterations. We could not obtain a reliable model under a single cool spot assumption, since the light curve asymmetries throughout the orbital phases show distinctive amounts of intensity drops, which is likely a consequence of different spot temperatures and/or radii. In the case of three cool spots, the solutions did not converge unless the temperature of the third spot would be equal to the photospheric temperature of the primary component. Consequently, we found a converged solution under the assumption of two cool spots on the primary component. The results we derived from this analysis are presented in Table~\ref{tab:table2}, while the model light curves obtained using the RV data from La Palma and the radial velocity curves from both datasets are shown in Figures~\ref{fig:figure5},~\ref{fig:rvfitlp}, and~\ref{fig:rvfitca}, respectively. The mass ratio (\textit{q}) obtained using the HERMES data from the simultaneous light curve and radial velocity data analysis is identical to that determined by \citet{Rucinski2002}, as well as the masses of both components, which are also consistent. The other absolute parameters (i.e. R$_{1,2}$, L$_{1,2}$) and the parameters related to system geometry (i.e. ${\rm \Omega_{1,2}}$, $i$) are also in accordance with those of given in the literature.

\begin{table}
	\centering
	\caption{Results from the simultaneous light and radial velocity curve analysis of SV Cam.}
	\label{tab:table2}
	\begin{tabular}{lccr} 
		\hline
		Stellar parameters & Solution 1 & Solution 2 \\
				  & (La Palma RVs) & (Calar Alto RVs) \\
		\hline
		$T{\rm _1} [K]$ & 6039 & 6039 \\
		$T{\rm _2} [K]$ & 4356(7) & 4364(10) \\
		$i \rm [^\circ]$ & 87.26(13) & 87.24(17) \\
		$q=M{\rm _2}/M{\rm _1}$ & 0.641(2) & 0.618(3) \\
        $K{\rm _1} [km/s]$ & 124.9(1.6) & 121.2(2.9) \\
        $K{\rm _2} [km/s]$ & 188.6(1.9) & 196.1(3.3) \\
		${\rm \Omega_{1}}$ & 3.58(1) & 3.54(1) \\
		${\rm \Omega_{2}}$ & 4.19(2) & 4.08(2) \\
		a [$R_{\odot}$] & 3.72(1) & 3.75(1) \\
		V$_{\gamma}$ [km/s] & -11.39(20) & -10.37(28) \\
		${\rm L_1/(L_1+L_2) [B]}$ & 0.967(6) & 0.966(6) \\
		${\rm L_1/(L_1+L_2) [V]}$ & 0.945(7) & 0.943(7) \\
		${\rm L_1/(L_1+L_2) [R]}$ & 0.924(6) & 0.922(7) \\
		${\rm L_1/(L_1+L_2) [I]}$ & 0.906(7) & 0.904(7) \\
		${\rm L_1/(L_1+L_2) [b]}$ & 0.956(7) & 0.955(7) \\
		${\rm L_1/(L_1+L_2) [y]}$ & 0.943(7) & 0.941(7) \\
		\hline
		\multicolumn{3}{|l|}{Spot parameters (on primary component)}\\
		\hline
		$\varphi_{1}$ [$^{\circ}$] & \multicolumn{2}{|c|}{41.3(1.2)} \\
		$\lambda_{1}$ [$^{\circ}$] & \multicolumn{2}{|c|}{294.8(1.1)} \\
		$\theta_{1}$ [$^{\circ}$] & \multicolumn{2}{|c|}{30.8(9)} \\
		$TF_{1}$ & \multicolumn{2}{|c|}{0.89(6)} \\
		$\varphi_{2}$ [$^{\circ}$] & \multicolumn{2}{|c|}{42.2(1.4)} \\
		$\lambda_{2}$ [$^{\circ}$] & \multicolumn{2}{|c|}{45.1(9)} \\
		$\theta_{2}$ [$^{\circ}$] & \multicolumn{2}{|c|}{30.1(1.2)} \\
		$TF_{2}$ & \multicolumn{2}{|c|}{0.90(7)} \\
		\hline
		\multicolumn{3}{|l|}{Absolute parameters}\\
		\hline
		M$_{1}$ [$M_{\odot}$] & 1.19(1) & 1.24(1) \\
		M$_{2}$ [$M_{\odot}$] & 0.76(2) & 0.77(2) \\
		L$_{1}$ [$L_{\odot}$] & 1.99(2) & 2.04(2) \\
		L$_{2}$ [$L_{\odot}$] & 0.21(2) & 0.24(2) \\
		R$_{1}$ [$R_{\odot}$] & 1.31(1) & 1.30(2) \\
		R$_{2}$ [$R_{\odot}$] & 0.81(2) & 0.81(2) \\
		R$_{1-pole}$ [$R_{\odot}$] & 1.25(1) & 1.27(2) \\
		R$_{2-pole}$ [$R_{\odot}$] & 0.79(2) & 0.80(2) \\
		logg$_{1}$ [cgs] & 4.29(3) & 4.30(3) \\
		logg$_{2}$ [cgs] & 4.52(3) & 4.51(3) \\
		\hline
		${\rm \Sigma(O-C)^2 [B]}$	& 0.048 & 0.047 \\
		${\rm \Sigma(O-C)^2 [V]}$	& 0.045 & 0.045 \\
		${\rm \Sigma(O-C)^2 [R]}$	& 0.110 & 0.109 \\
		${\rm \Sigma(O-C)^2 [I]}$	& 0.250 & 0.251 \\
		${\rm \Sigma(O-C)^2 [b]}$	& 0.061 & 0.060 \\
		${\rm \Sigma(O-C)^2 [y]}$	& 0.055 & 0.055 \\
		\hline
	\end{tabular}
\begin{threeparttable}
   \begin{tablenotes}
      \small
      \item $\varphi_{i}$ - colatitude of spot, $\lambda_{i}$ - longitude of spot, $\theta_{i}$ - angular radius of spot, $TF_{i}$ - temperature factor of spot. Formal errors from the PHOEBE code were given in parentheses. Errors of spot parameters were obtained during the solution using La Palma RVs.
    \end{tablenotes}
\end{threeparttable}
\end{table}

\section{Doppler Imaging}

In order to map the primary component of SV Cam, we used the Doppler imaging code DoTS \citep{Cameron1997}, which, based on the Maximum Entropy Method (MEM), finds the best-fitting spot distribution across the stellar surface. The code can also be used to determine the system parameters accurately, by performing maximum entropy regularized iterations to find which set of parameters yields the best fit to the line profiles. The parameter optimization is very critical in that sense, as the inaccurate determination of system parameters can lead to artifacts in the surface maps \citep[e.g.][]{Unruh1996}. Moreover, compared to single stars, Doppler imaging of binary systems requires many more parameters, making the accurate system parameter determination crucial.

Based on the two-temperature model to represent photosphere(s) and spot(s), DoTS needs lookup tables generated using the spectra of template stars to calculate the intensity contributions of each element across the stellar surface. In this context, the spectra of HD~143761 (G0V) and HD~4628 (K2.5V), obtained during both Calar Alto and La Palma observing runs, are used to represent the photospheres of the primary and secondary components. Assuming the spot temperature to be 1000 K lower ($\sim$ 5000 K) than the photosphere of the primary component, as follows from the studies by \citet{Zboril2004} and \citet{Hempelmann1997}, we used the spectra of HD~22049 (K2V) from both runs to represent the spectrum of the star spot(s). We determined the linear limb darkening coefficients using the tables by \citet{Hamme1993}, taking into account the effective temperatures of both components.

\begin{figure}
\centering
	\includegraphics[width=\columnwidth]{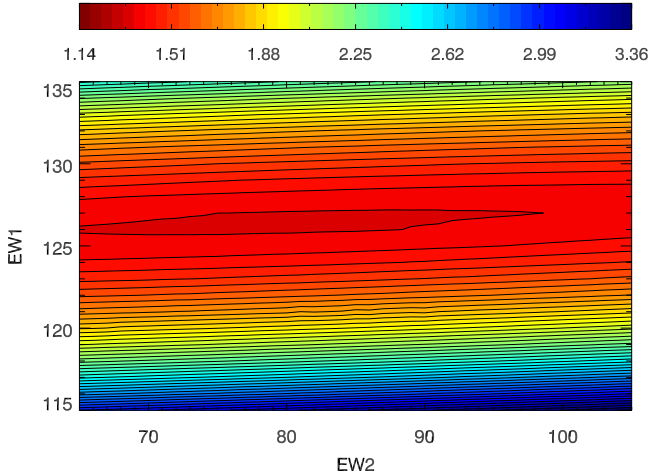}
 	\caption{Two-dimensional grid search for the optimization of $EW_{1,2}$ parameters.}
    \label{fig:ewcont}
\end{figure}

\begin{figure}
\centering
	\includegraphics[width=\columnwidth]{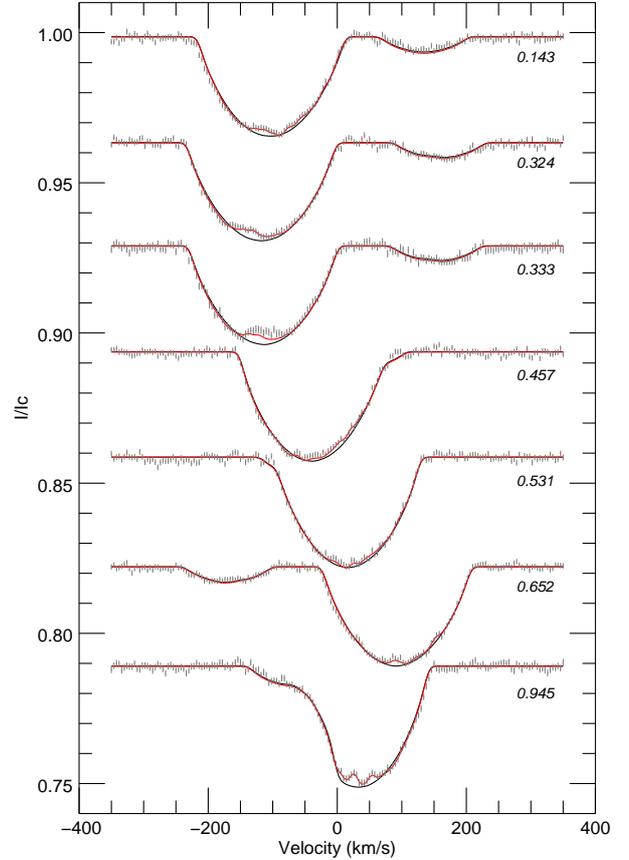}
 	\caption{Phase-ordered LSD profiles together with the error bars for Calar Alto dataset. Black solid lines represent the synthetic velocity profiles generated using the system parameters, while the red solid lines show the maximum entropy regularized models of SV Cam.}
    \label{fig:lsdfitca}
\end{figure}

\begin{figure*}
\centering
	\includegraphics[width=20pc,height=38pc,angle=270]{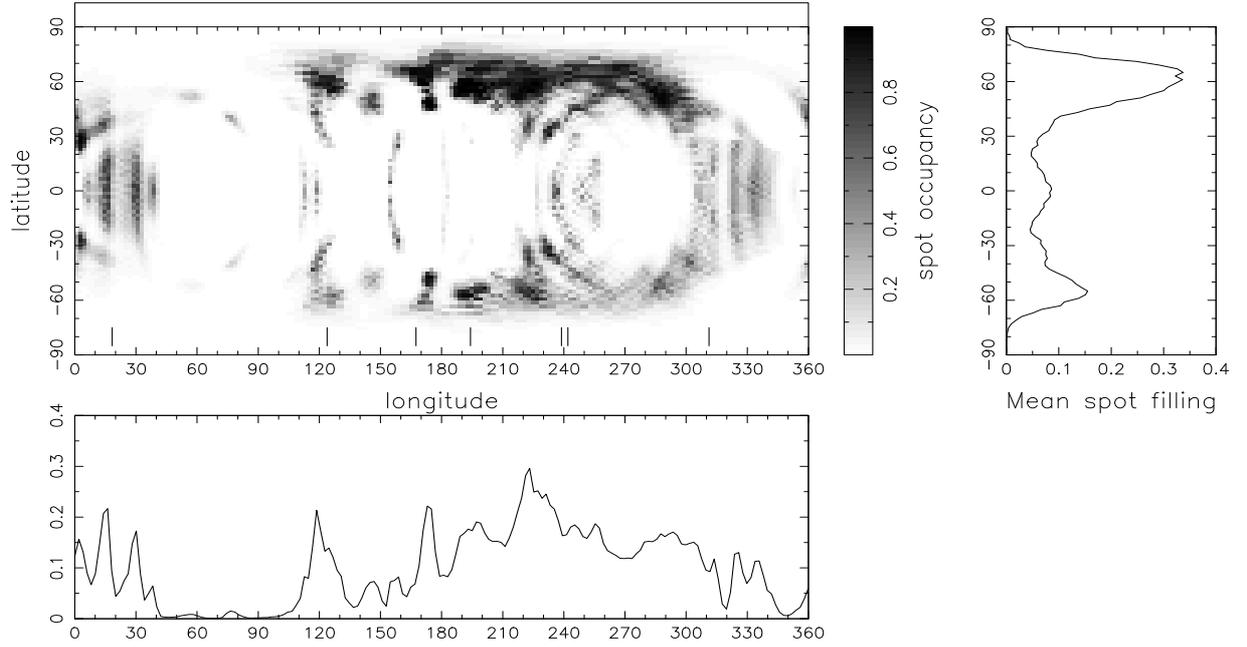}
 	\caption{Mercator projection of reconstructed image for the primary component of SV Cam obtained using the Calar Alto dataset that are acquired between 19 and 31 October, 2015. Right panel shows the latitudinal spot filling factor, while the lower left panel shows the longitudinal correspondence.}
    \label{fig:camap}
\end{figure*}

\begin{figure*}
\centering
	\includegraphics[trim=1 5 1 30,clip,width=31pc,height=29.5pc]{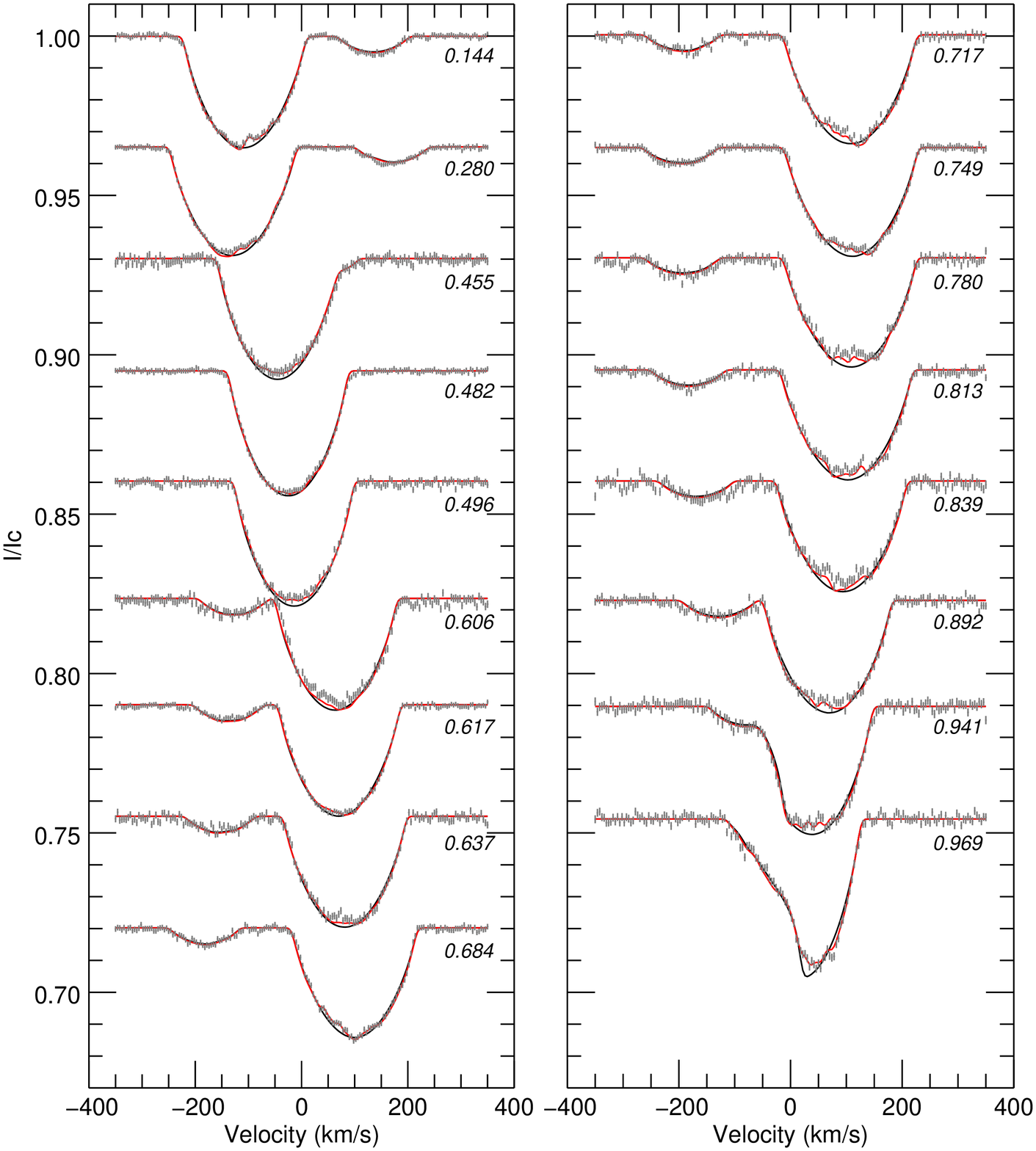}
 	\caption{Phase-ordered spectra for the La Palma dataset. Error bars and synthetic profiles are plotted as for the Calar Alto data in Figure \ref{fig:lsdfitca}.}
    \label{fig:lsdfitlap}
\end{figure*}

\begin{figure*}
\centering
	\includegraphics[width=20pc,height=38pc,angle=270]{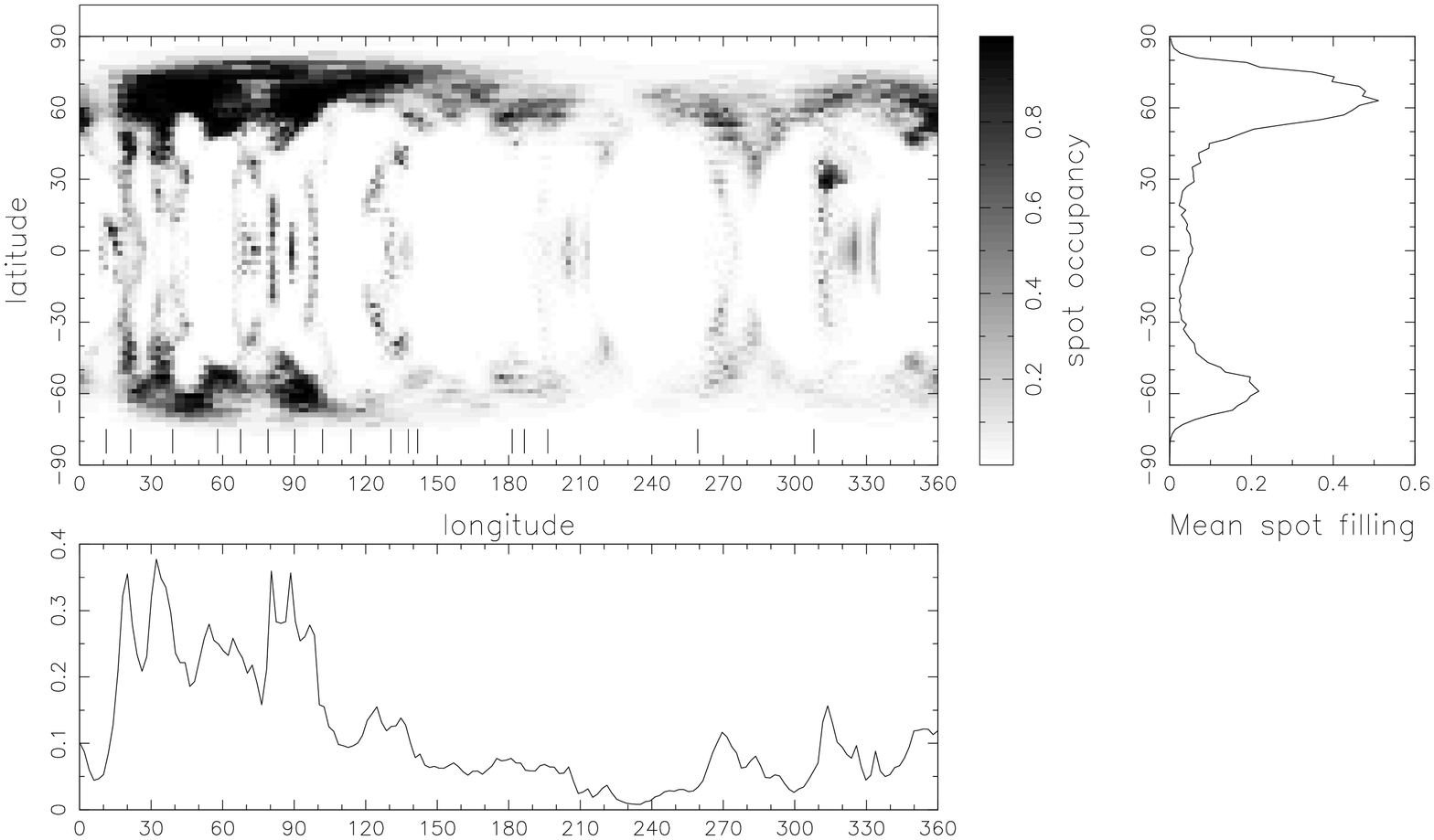}
 	\caption{Same as Figure \ref{fig:camap} but for La Palma dataset obtained between 12 and 18 January, 2017.}
    \label{fig:lpmap}
\end{figure*}

Under the assumption of two cool spots, the simultaneous analysis of high-quality \textit{BVRIby} light curves together with the radial velocities enabled us to estimate the system parameters accurately. During the surface reconstruction via DoTS, we used the set of parameters obtained from simultaneous LC + RV solutions from La Palma dataset. These RV measurements are more numerous and therefore the obtained results are more reliable, especially for the system orbital parameters. As can be seen from Table~\ref{tab:table2}, center-of-mass velocities (V$_{\gamma}$) reckoned by the solutions of the two RV curves obtained with HERMES and CAFE spectrographs more than one year apart from each other have $\sim$\,1\,km/s difference.
In addition, the orbital period variation of the system shows a cyclic variation that is attributed to the existence of an additional body, in the literature. This V$_{\gamma}$ variation could be due to a third body, but we cannot exclude a systematic effect related to the different instruments. However, this difference in V$_{\gamma}$ was taken into account during the surface reconstruction performed for both datasets. PHOEBE gives the arithmetic mean of r$_{pole}$, r$_{side}$ and r$_{back}$ parameters, whereas DoTS requires the polar radii as the input parameter. The errors in Table~\ref{tab:table2} are the formal errors given by the PHOEBE code.

The EW parameter in the DoTS code controls the strength of the synthetic LSD lines that are generated using the spectra of template stars, which represent the local intensity profiles of the photosphere(s) and spot(s). This is adjusted to optimally fit the entire dataset as part of the parameter optimisation process. Therefore, the accurate determination of line $EW_{1,2}$ parameters are also critical, as it affects both the widths and depths of the modelled line profiles. This is particularly important in preventing the artificial high latitude spots during the surface reconstruction. In order to determine the optimum values of $EW_{1,2}$, we carried out a 2-dimensional grid search separately for both datasets, using DoTS for the optimization of $EW_{1,2}$. The resulting values of $EW_{1,2}$ optimization for the Calar Alto dataset are shown in Figure~\ref{fig:ewcont}. During surface reconstructions, we used the TEST statistics option of DoTS to determine the degree of convergence for maximum entropy as the stopping criteria \citep[see][for details]{Senavci2011}. The datasets and the best-fit models are shown in Figures \ref{fig:lsdfitca} and \ref{fig:lsdfitlap}, while the corresponding maps are in Figures \ref{fig:camap} and \ref{fig:lpmap}, respectively. Both sets of maps show a symmetric distribution of spots, which can be attributed to the inability of DI techniques to determine on which the hemisphere the spot(s) are located at, in systems with inclination angles approaching 90$^\circ$.

\section{Photometric Mapping}
\label{sec:photmap}

\begin{figure}
\centering
	\includegraphics[trim=1 20 1 50,clip,width=\columnwidth]{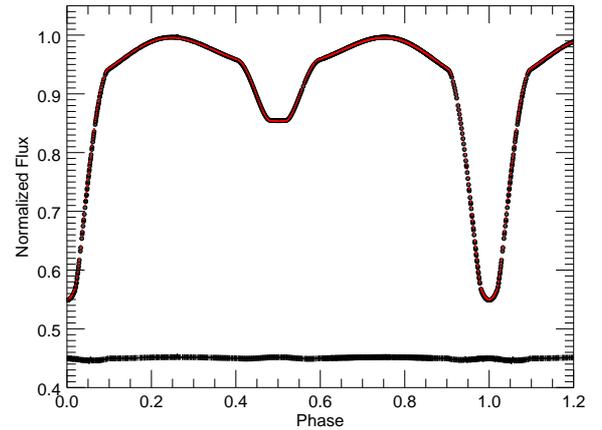}
 	\caption{Comparison of unspotted R-band light curves of SV Cam obtained using DoTS (black circles) and PHOEBE (red circles) together with the residuals.}
    \label{fig:lcphbdots}
\end{figure}

Prior to mapping our photometric light curves using DoTS we must first simulate the unspotted light curve. To do this, we carried out the following steps. We chose the R-band light curve of the system, since the central wavelength of our spectral range is close to that band. We then generated an unspotted light curve with DoTS, using exactly the same parameters from PHOEBE solution and local intensity models based on the PHOENIX models. A comparison of unspotted synthetic light curves generated using both DoTS and PHOEBE codes is shown in Figure \ref{fig:lcphbdots}. The ${\rm \Sigma(O-C)^2}$ value of 0.0025 obtained from that comparison clearly shows that the output from both codes is in agreement.

\begin{figure}
\centering
	\includegraphics[trim=1 20 1 50,clip,width=\columnwidth]{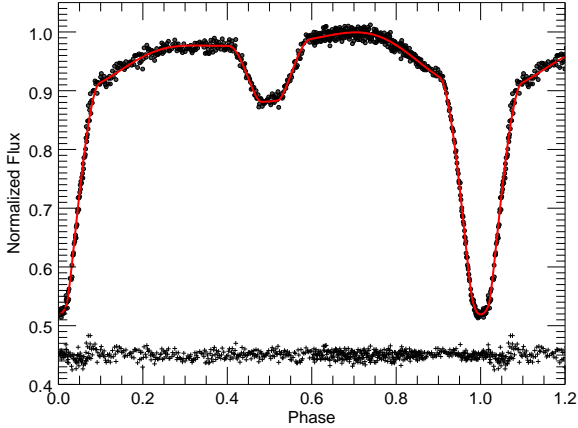}
 	\caption{The best-fit (red solid lines) and residuals (plus signs) for the R-band data of SV Cam obtained using the photometric mapping mode of the code DoTS. Here the ${\rm \Sigma(O-C)^2}$ is 0.058. The resultant map (Fig.~\ref{fig:phomap}) shows two predominant spot features centered at around longitudes $\sim$50$^{\circ}$ and 300$^{\circ}$.}
    \label{fig:dotslcfits}
\end{figure}

\begin{figure}
\centering
	\includegraphics[trim=1 1 1 40,clip,width=\columnwidth]{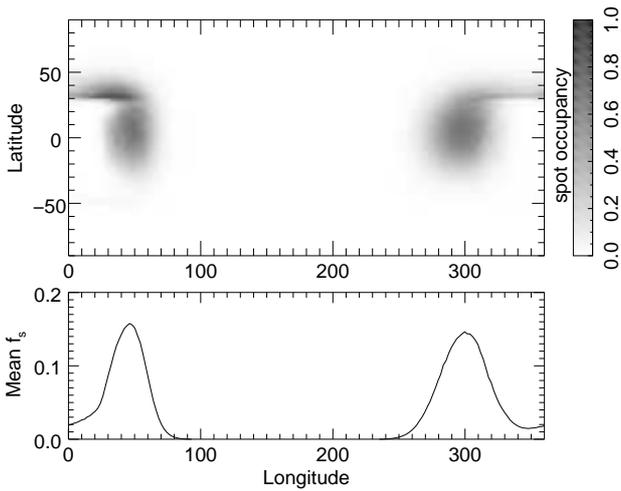}
 	\caption{Mercator projection (upper panel) and longitudinal spot filling factor distribution (lower panel) of reconstructed image for the primary component of SV Cam obtained using the R-band data of the system acquired between 31 December 2015 and 01 January 2016. Note that no latitude information can be obtained from photometry using only one band.}
    \label{fig:phomap}
\end{figure}

After testing the consistency of unspotted system configuration, we carried out the photometric mapping for the primary component of SV Cam. The best fit to the R-band light curve of the system and the resulting map for the primary component are plotted in Figures \ref{fig:dotslcfits} and \ref{fig:phomap}, respectively. Further tests and a discussion of the map will be given in Sect.~\ref{sssec:photest}.

\section{Results and Discussion}

We have investigated the surface inhomogeneities of the primary component of the RS CVn type eclipsing binary SV Cam via the code DoTS, using high resolution time-series spectra obtained from Calar Alto and La Palma observatories. The high quality multi-band photometric data, taken in two consecutive nights, also allowed us to determine the system parameters accurately, with the help of the light and radial velocity curve analysis performed simultaneously. We also obtained the surface map of the primary component using the R-band light curve of the system, this time using the photometric mapping mode of DoTS.
The Calar Alto data includes only 7 time-series spectra mostly distributed between orbital phases 0.0 - 0.5 that corresponds to the longitudes 360$^{\circ}$ to 180$^{\circ}$ (note that phase runs in reverse to longitude).  The La Palma data including 17 time-series spectra on the other hand, have better phase coverage mainly concentrated between longitudes 0$^{\circ}$ and 200$^{\circ}$. However, we have only 2 spectra distributed between longitudes 200$^{\circ}$ and 360$^{\circ}$. The tests showed that surface reconstructions with such poor data sampling, especially like the Calar Alto data in our case, mostly carry reliable information about the distribution of large-scale spots and show smearing or absence of small-scale starspot features in the surface maps \citep{Xiang2015}.

\subsection{Star-spot distributions}

\subsubsection{Doppler images}

The surface map obtained using Calar Alto data (Figure \ref{fig:camap}) shows high-latitude spot features extended to lower latitudes and predominantly located between longitudes 150$^{\circ}$ and 300$^{\circ}$, with a maximum longitudinal spot filling factor value of $\sim$0.3. There is a spotless zone between longitudes $\sim$40$^{\circ}$ and 100$^{\circ}$, due to the deficiency and uneven distribution of data throughout the orbital cycle. The surface map from La Palma data (Figure \ref{fig:lpmap}) shows predominant high-latitude spot features between longitudes 0$^{\circ}$ and 150$^{\circ}$ with a slightly higher filling factor than that of the map obtained using Calar Alto data. It is not possible to make interpretations on the migration of spot features, due to the inhomogeneous distribution of spectral data and the large interval between the two observing runs ($\sim$ 14 months). What is clear is that there is heavy spot coverage where there is good phase coverage in both maps. 
This can represent the actual longitudinal distribution of spots, 
which is similar to the those obtained by \citet{Hempelmann1997}, who also 
used a MEM based DI code, but with more evenly sampled phases throughout 
the orbital period. 
Alternatively, the actual distribution can be more homogeneous 
(axisymmetric), and the longitudinal profile we obtained was thus 
affected by the phase sampling.

The overall distribution of high-latitude spot features, as well as the extensions to the lower latitudes are clear from our maps. This is expected in such active rapid rotators, due to the enhanced Coriolis force acting on rising flux tubes (\citealt{ss92}, \citealt{holzwarth03b}, \citealt{Isik2011}). \citet{Jeffers2005} revealed the existence of a polar spot on the primary component using precise spectrophotometric HST data. The presence of a polar spot on the surface of an RS CVn type binary $\zeta$ And was directly imaged by \citet{Roettenbacher2016}, using long-baseline infrared interferometry. Besides, most of the RS CVn type binaries are known to have polar caps \citep{Strassmeier2009}. However, most of such systems have lower orbital inclinations than that of SV Cam. We will therefore test the reliability of the latitudinal distributions of our maps, which do not show polar spots, in Sect.~\ref{sssec:inctest}.

\subsubsection{Numerical tests on phase smearing}
\label{sssec:smeartest}
\begin{figure}
\centering
\begin{tabular}{c}
\includegraphics[width=18.0pc]{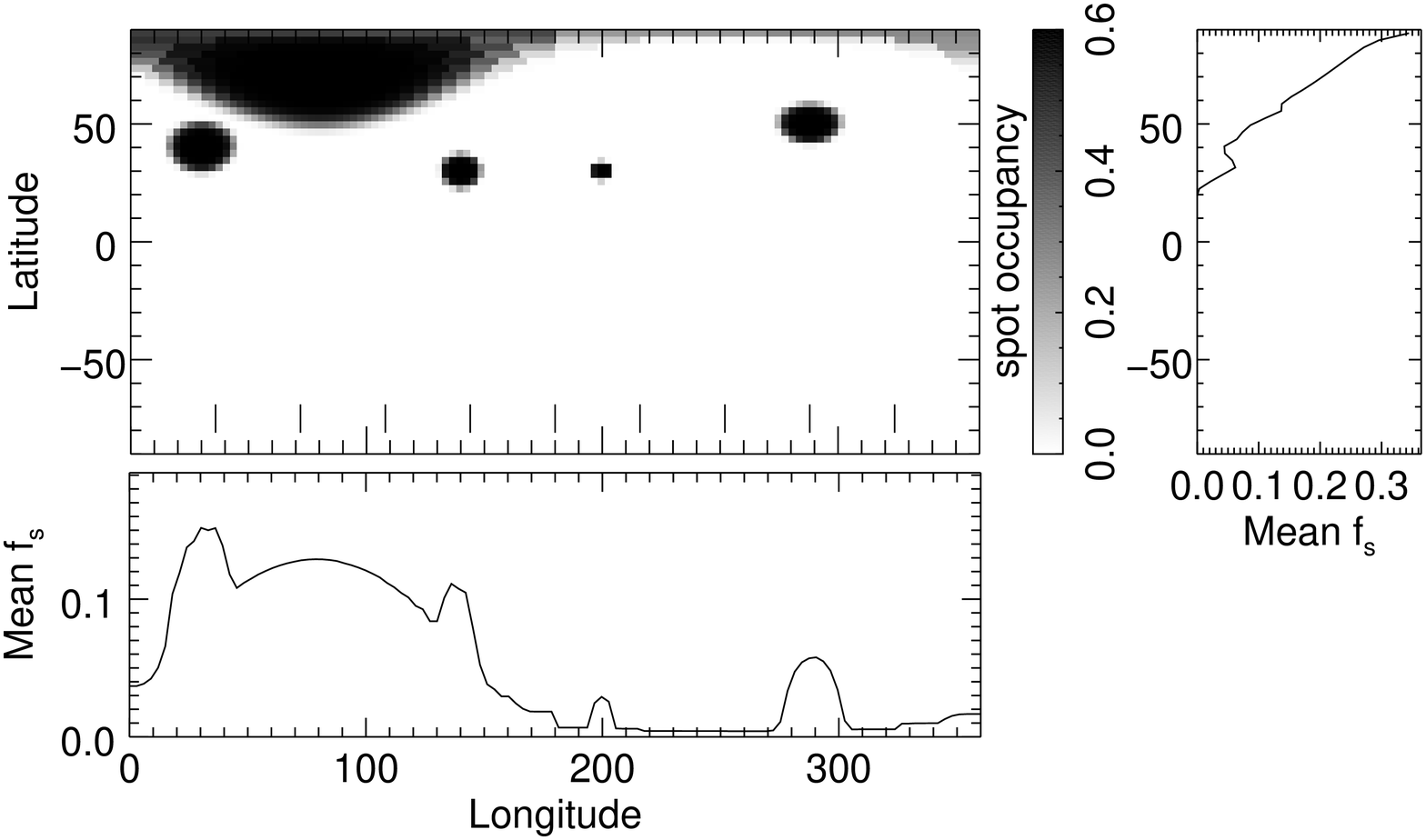}
\end{tabular}
\begin{tabular}{c}
\includegraphics[width=18.0pc]{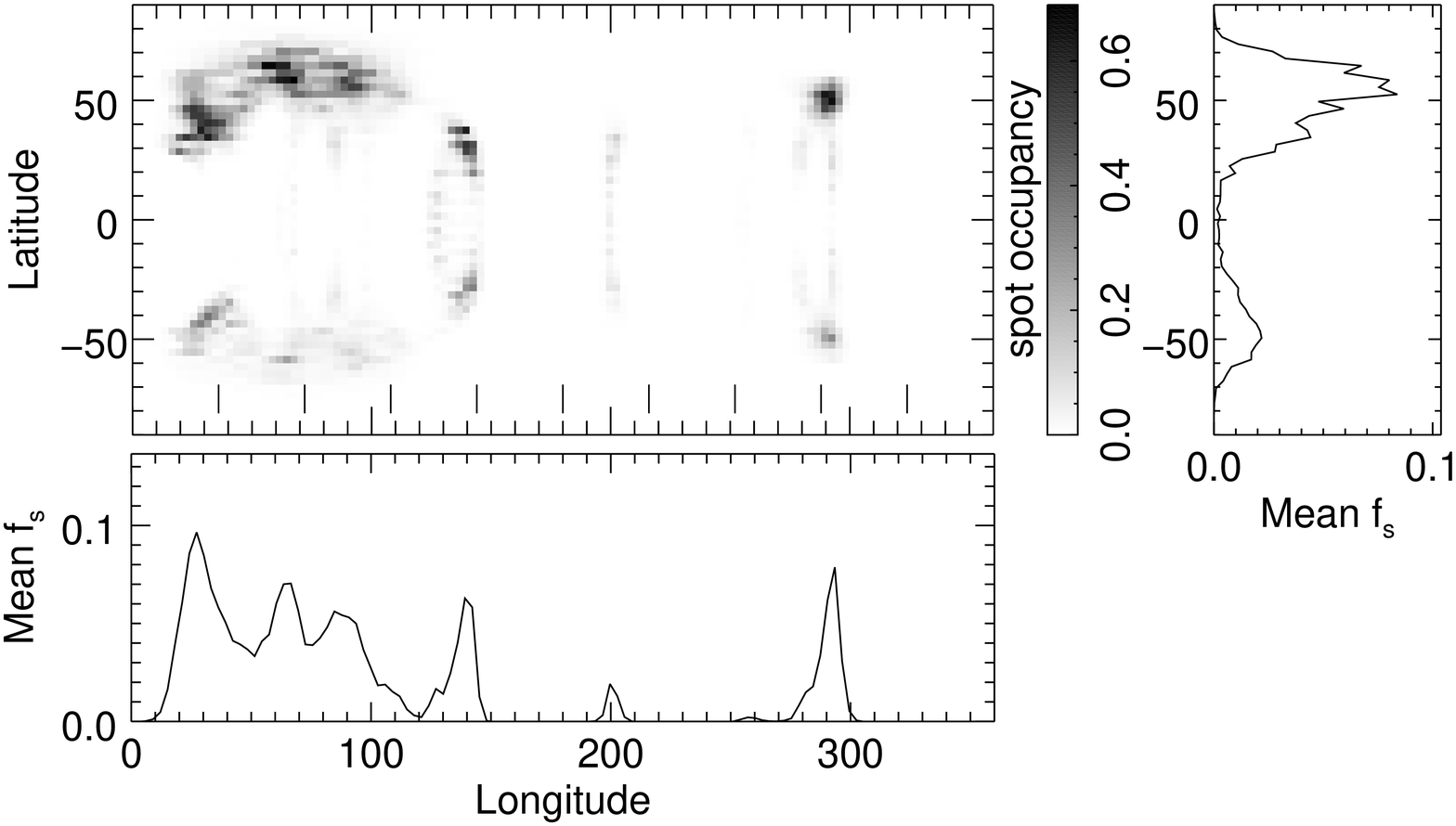}
\end{tabular}
\begin{tabular}{c}
\includegraphics[width=18.0pc]{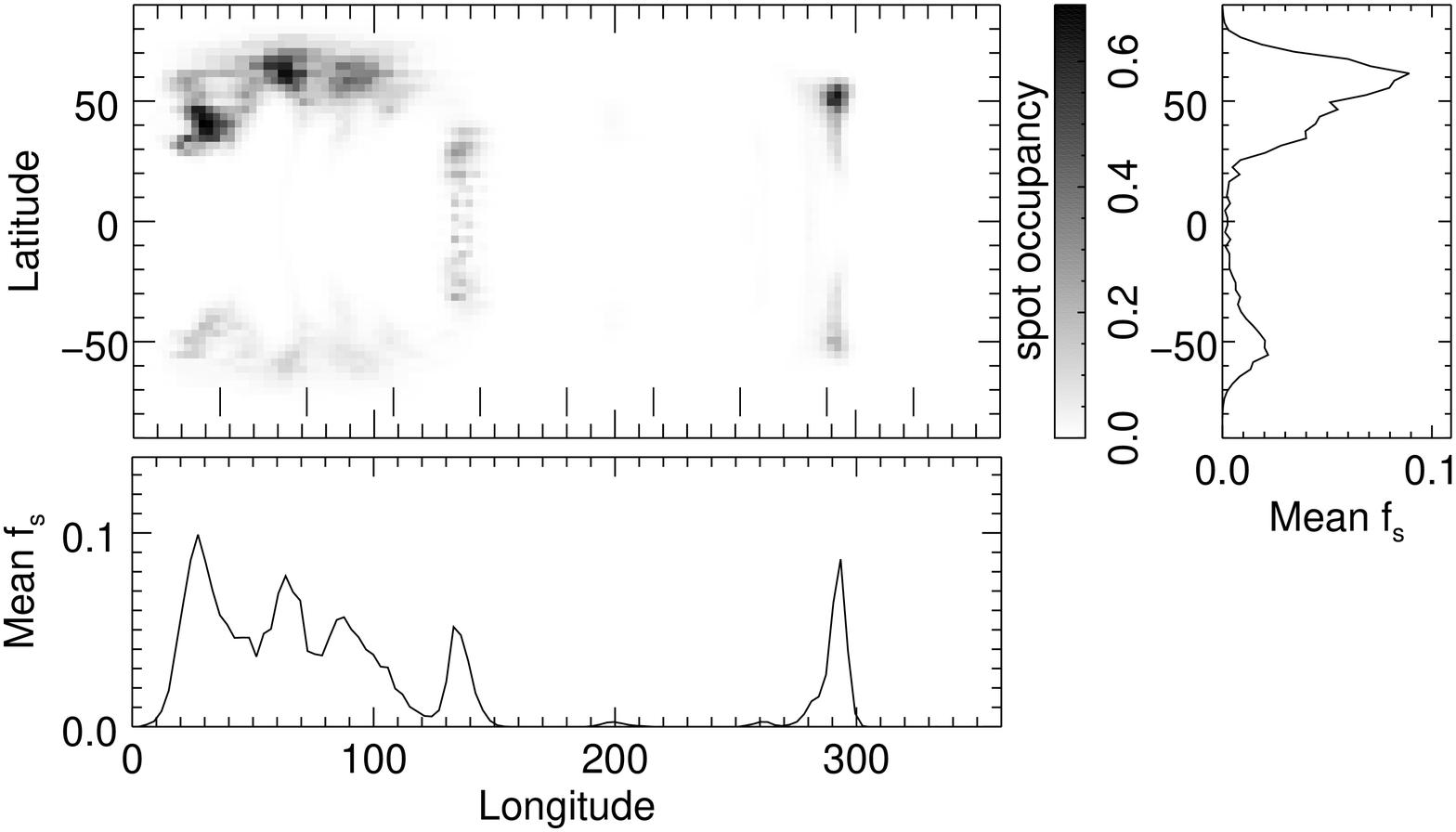}
\end{tabular}
\caption{Reconstructions for the input spot map (upper panel); S1.0 (middle panel) and S3.5 (lower panel).}
\label{fig:testsmear}
\end{figure}

The CAFE data of SV Cam was obtained using 1500 seconds of exposure time, while 1500, 1600 and 1800 seconds of exposure times were used during the HERMES observing run. These exposure times correspond to 2.9\%, 3.1\% and 3.5\% of the orbital period of SV Cam, respectively, which cause the well-known phase smearing phenomenon. The smearing produced by the maximum exposure time of 1800 seconds is about 0.035 in units of the orbital period, $\sim$ 12$^{\circ}$ in longitude, or $\sim$ 15 km/s in radial velocity. It causes the blurring of the LSD profiles and hence affects the resultant maps, preventing the reconstruction of smaller surface features that can be resolved via shorter exposure times. It is possible to take into account the phase smearing effect within the DoTS code, under a reasonable number of profiles as well as the number of data points for each profile, depending on the available computational power and memory limitations. We have not carried out a direct high-performance reconstruction of the Doppler images by considering the exposure times. Instead, we have tested the effects of smearing, by generating artificial profiles resulting from five spots on the primary component with different locations and sizes (between 4$^{\circ}$ and 20$^{\circ}$), using the same system parameters of SV Cam. The artificial LSD profiles were generated with 0.1 phase intervals. To mimic the smearing levels of 1\% and 3.5\% of the orbital period, the artificial data were produced using 500 and 1800 seconds of exposure times, respectively. We call these two cases S1.0, corresponding to the conventional theoretical limit to avoid any measurable effects of smearing, and S3.5, to represent our longest exposure time. The resulting reconstructions are shown in Figure \ref{fig:testsmear}.

The highest-latitude spot on both S1.0 and S3.5 maps is reconstructed at lower latitudes, as a consequence of high orbital inclination of the system, which is discussed in Section \ref{sssec:inctest}, in detail. The reconstructed spots on the S3.5 map are blurred compared to those on S1.0. The smallest spot, which has a radius of 4$^{\circ}$ and located at 200$^{\circ}$ longitude, is resolved on S1.0 and disappears on S3.5. Therefore, our exposure times do not lead to considerable uncertainties for relatively large spots in Figs. \ref{fig:camap} and \ref{fig:lpmap}, whereas features with sizes smaller than about 12$^{\circ}$ (corresponding to our longest exposure time) can be artefacts of the inversion procedure.

\subsubsection{Numerical tests on the orbital inclination}
\label{sssec:inctest}
\begin{figure}
\centering
\begin{tabular}{c}
\includegraphics[width=18pc]{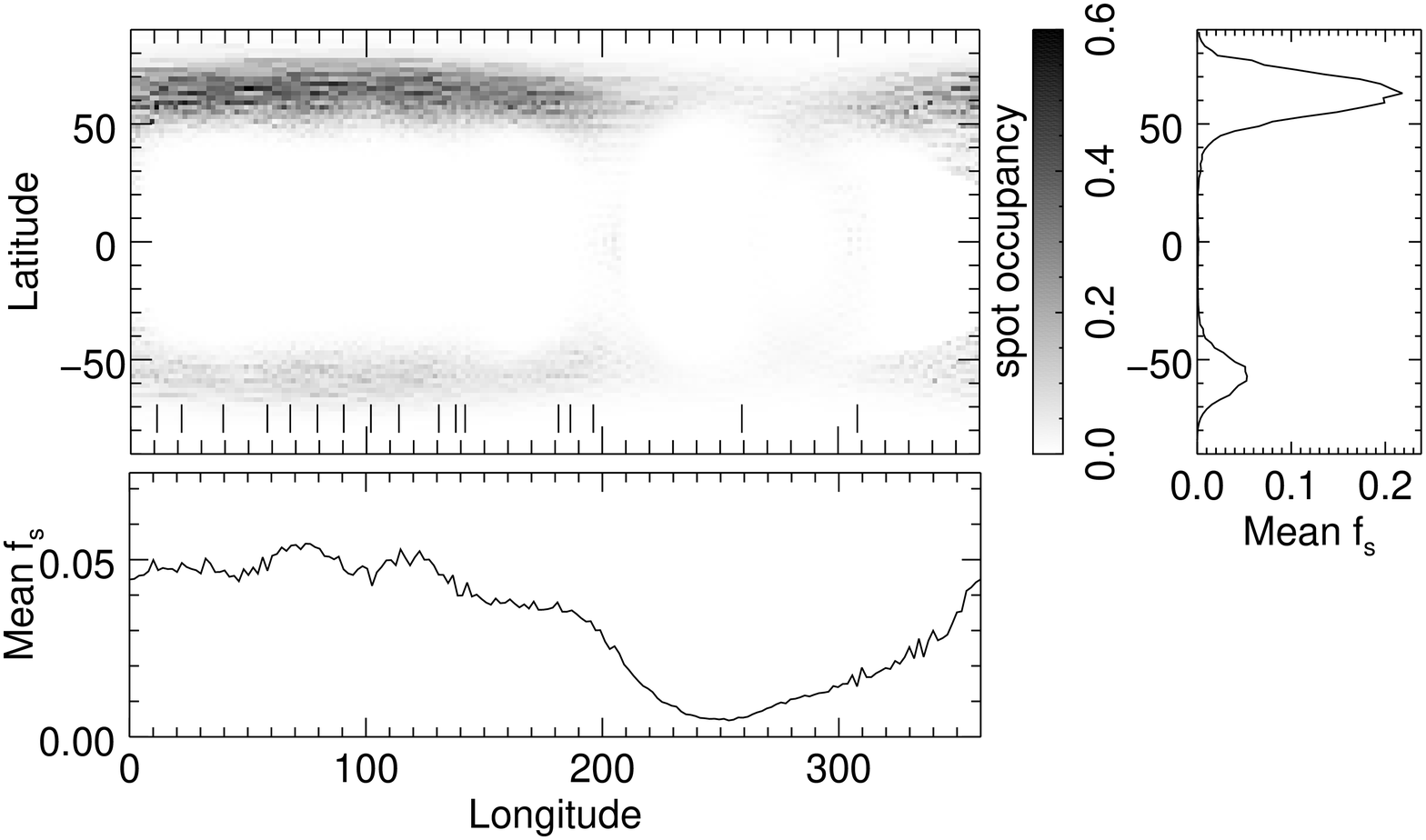}
\end{tabular}
\begin{tabular}{c}
\includegraphics[width=18pc]{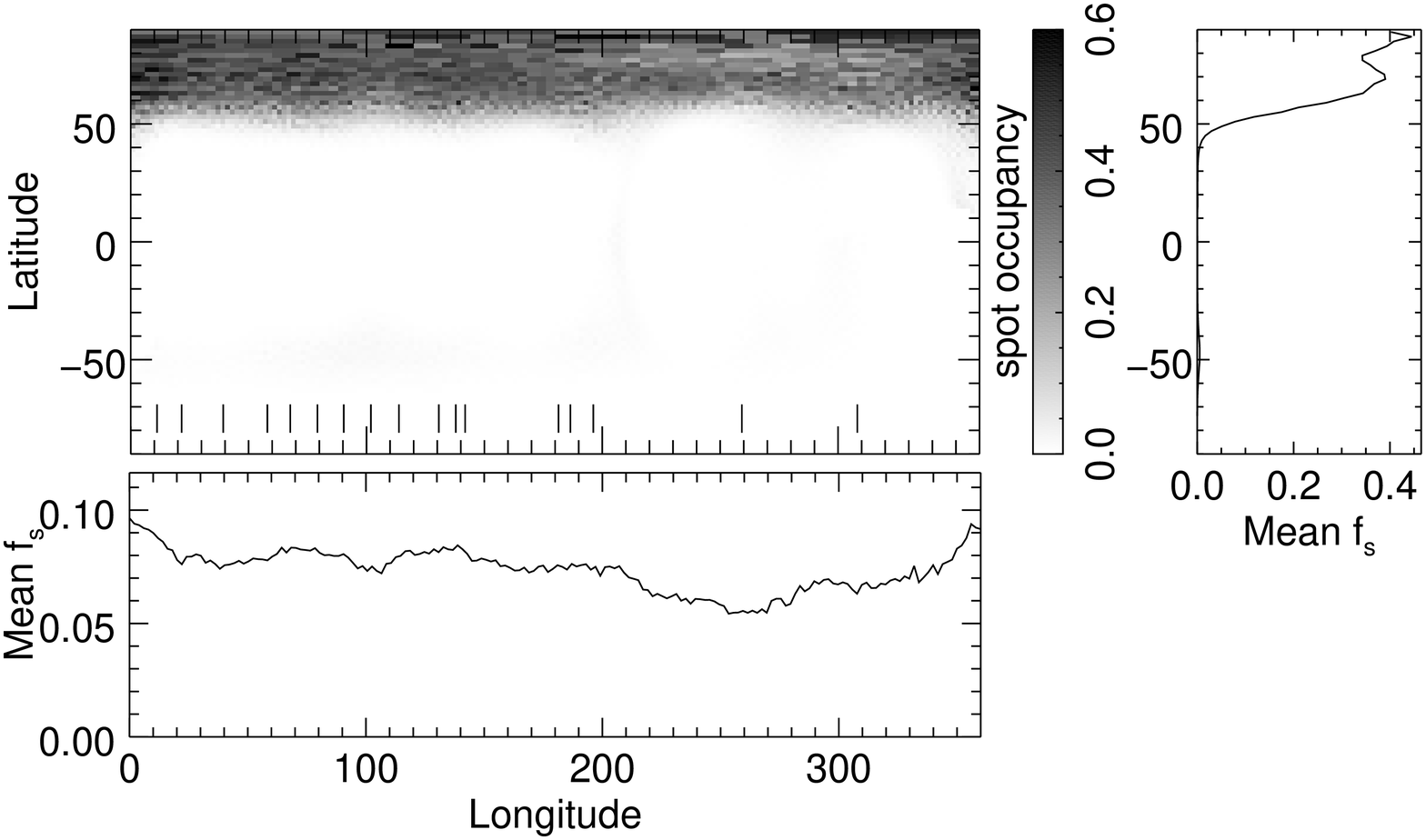}
\end{tabular}
\begin{tabular}{c}
\includegraphics[width=18pc]{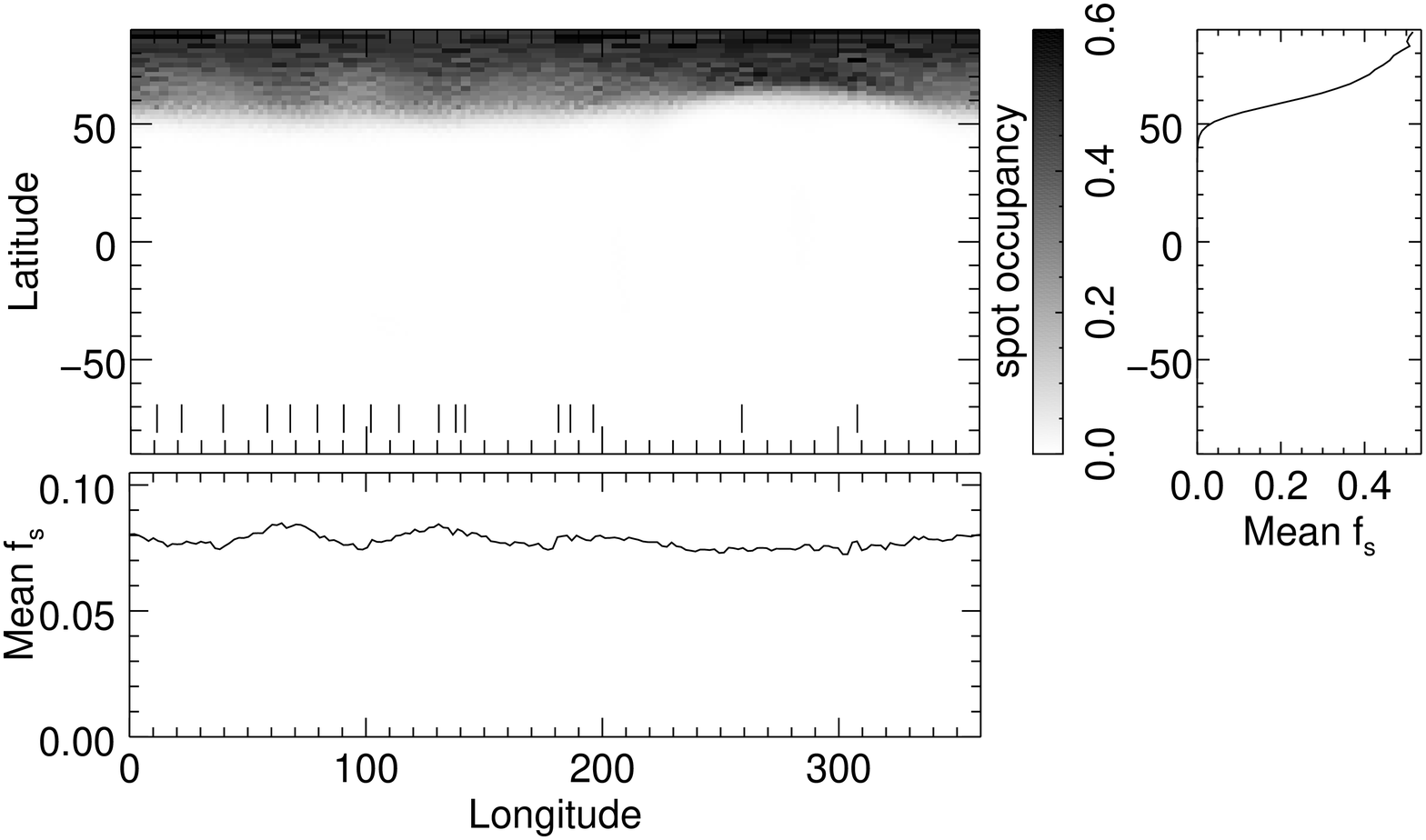}
\end{tabular}
\caption{Test results concerning the reconstruction of a polar spot under the orbital inclinations i=87.2$^{\circ}$ (top panel), 80$^{\circ}$ (middle panel) and 70$^{\circ}$(bottom panel).}
\label{fig:testpole}
\end{figure}

We performed simulations in order to test the ability of our surface reconstructions to recover polar spots at different inclination angles approaching 90 degrees. In this context, using exactly the same system parameters of SV Cam, except for the orbital inclination, we simulated a polar spot with a radius of 30$^{\circ}$ on the primary component and generated synthetic profiles for three chosen orbital inclinations:  $i$=87.2$^{\circ}$ (the value obtained from simultaneous LC and RV analysis), 80$^{\circ}$ and 70$^{\circ}$. Synthetic profiles were generated using the data sampling of La Palma observations to test the effects of phase coverage. The results are shown in Figure \ref{fig:testpole}. It is clear from the top panel of Figure \ref{fig:testpole} that the resultant map shows a high latitude spot located at latitude of  $\sim$60$^{\circ}$  together with a symmetric counterpart on the southern hemisphere. The surface map shown in the middle panel of Figure \ref{fig:testpole}, shows a ``polar like" spot, a feature having a double-peaked filling factor ($f_s$) distribution between $\sim$70$^{\circ}$ and 90$^{\circ}$, while the lower panel clearly exhibits a polar spot with spot coverage peaking at 90$^{\circ}$. Furthermore, the effect of poor phase sampling (between $\sim$210$^{\circ}$ - 360$^{\circ}$) is obviously seen in all surface maps. These results indicate that the existence of polar caps has to be taken into account with some care as one approaches 90$^{\circ}$, since this may be due to a maximum entropy effect whereby it is less penalising to add a high latitude band rather than a complete polar spot. 
Therefore, while the mid-low latitude structure is reliable, we cannot rule out a polar spot instead of the high latitude spots seen on our maps of the primary component. It is also a known phenomenon that the spots near the equator have smearing and vertical elongation effects as a consequence of poor latitude discrimination of Doppler imaging around the equator, especially for stars with high inclinations and poor data sampling \citep{Cameron1994}. Such features are visible in the surface map obtained from Calar Alto data at around longitudes 0$^{\circ}$-30$^{\circ}$ and 300$^{\circ}$-330$^{\circ}$, while similar features are also visible in the surface map obtained using La Palma data between longitudes 210$^{\circ}$ - 360$^{\circ}$. In the surface map based on the La Palma dataset, where we have good sampling between longitudes 0$^{\circ}$ and 200$^{\circ}$, low to mid latitude spots together with the high latitude (or polar) ones are obvious. The two images separated by about 1 year are qualitatively similar, with a strong high-latitude (or near-polar) spot complex and scattered low-latitude spots with a smaller filling factor that spread over the entire surface. Polar spots on
active cool stars can live very long (\citealt{Strassmeier2009}, \citealt{Hussain2002}) possibly owing to emergence of bipolar magnetic regions at very high latitudes and with large tilt angles \citep{Isik2007}, but since the polarity information is not available here, we cannot discuss the details about magnetic structure and evolution of these features. We also performed additional simulations to determine the orbital inclination limit that causes the reconstruction of high latitude spots rather than polar ones as well as the vertical elongation, for SV Cam configuration. The results showed that the input polar spot becomes a high latitude band and the vertical elongation begins when the orbital inclination is \textit{i} = 85$^{\circ}$.

\subsubsection{Light curve inversions and numerical tests on latitudinal dependence}
\label{sssec:photest}

Figure \ref{fig:phomap} shows the surface map obtained using R-band light curve of the system with two main spot features centered at around longitudes $\sim$50$^{\circ}$ and 300$^{\circ}$, which are also in accordance with the location of spots estimated via PHOEBE (Sect.~\ref{sec:photmap}). 
Numerical tests using the DoTS code with the parameters of RS CVn-type binaries have shown that the longitudinal accuracy of the reconstructions 
is sensitive to the accuracy of the system parameters such as effective 
temperatures and radii of components \citep{Jeffers2005,ozavci18}. We note that in our R-band light curve inversion, the system parameters of SV Cam are preset to values obtained from the simultaneous light and radial velocity solution, so we do not expect in our inversion persistent, spurious spots near quadrature phases related to random background spots \citep[see][Sect.~3.2]{ozavci18}.

The symmetry of spots with respect to the equator is again a consequence of a high orbital inclination. Normally, it is not possible to obtain latitudinal information from photometry, because a light curve represents one-dimensional time series, hence the resulting map contains only the longitudinal spot distribution information \citep{Berdyugina2005}. However, the resultant map from our light curve inversion shows latitudinal structuring, especially around latitude 30$^{\circ}$.

\begin{figure}
\centering
\begin{tabular}{c}
\includegraphics[width=18.0pc]{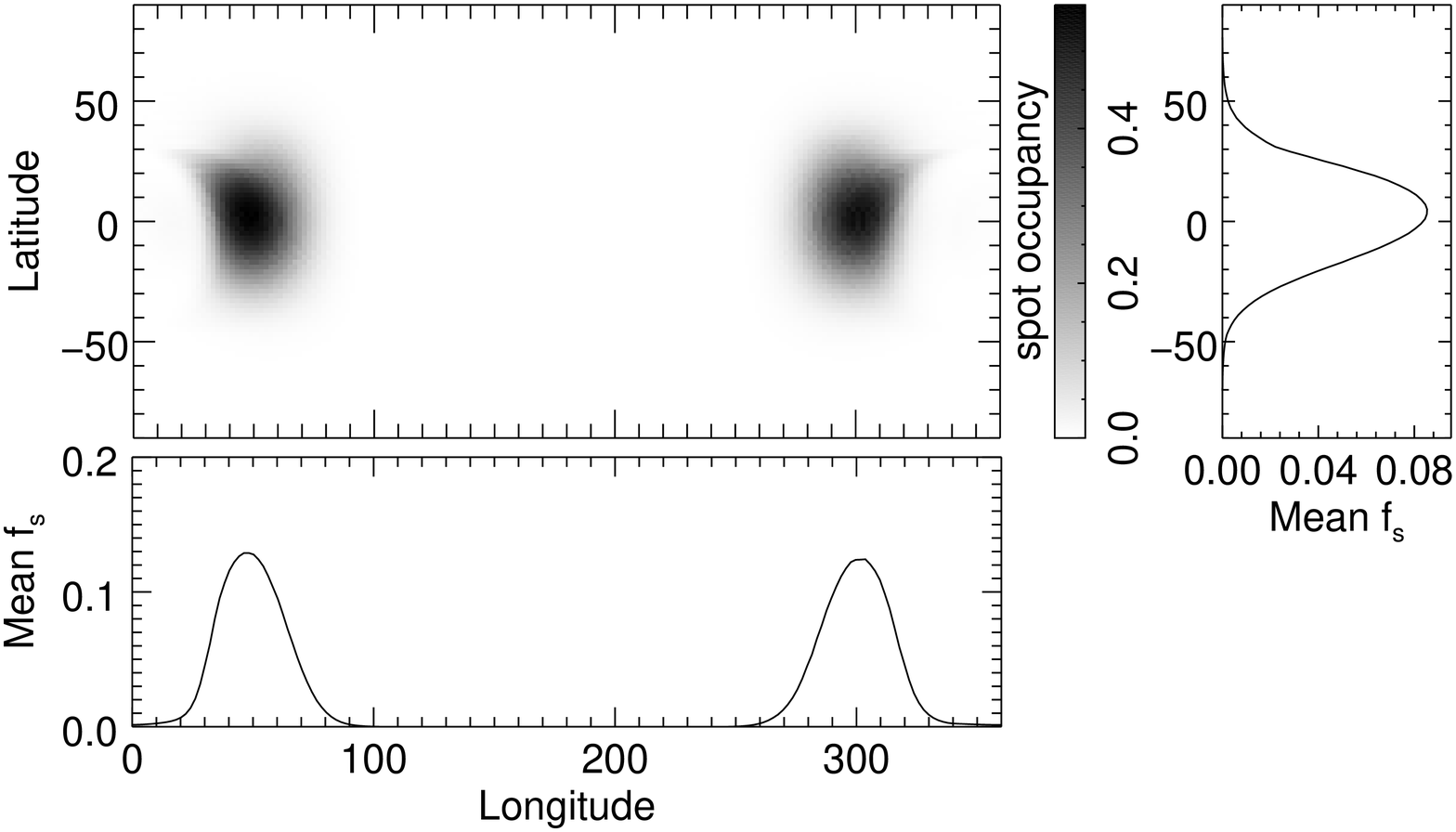}
\end{tabular}
\begin{tabular}{c}
\includegraphics[width=18.0pc]{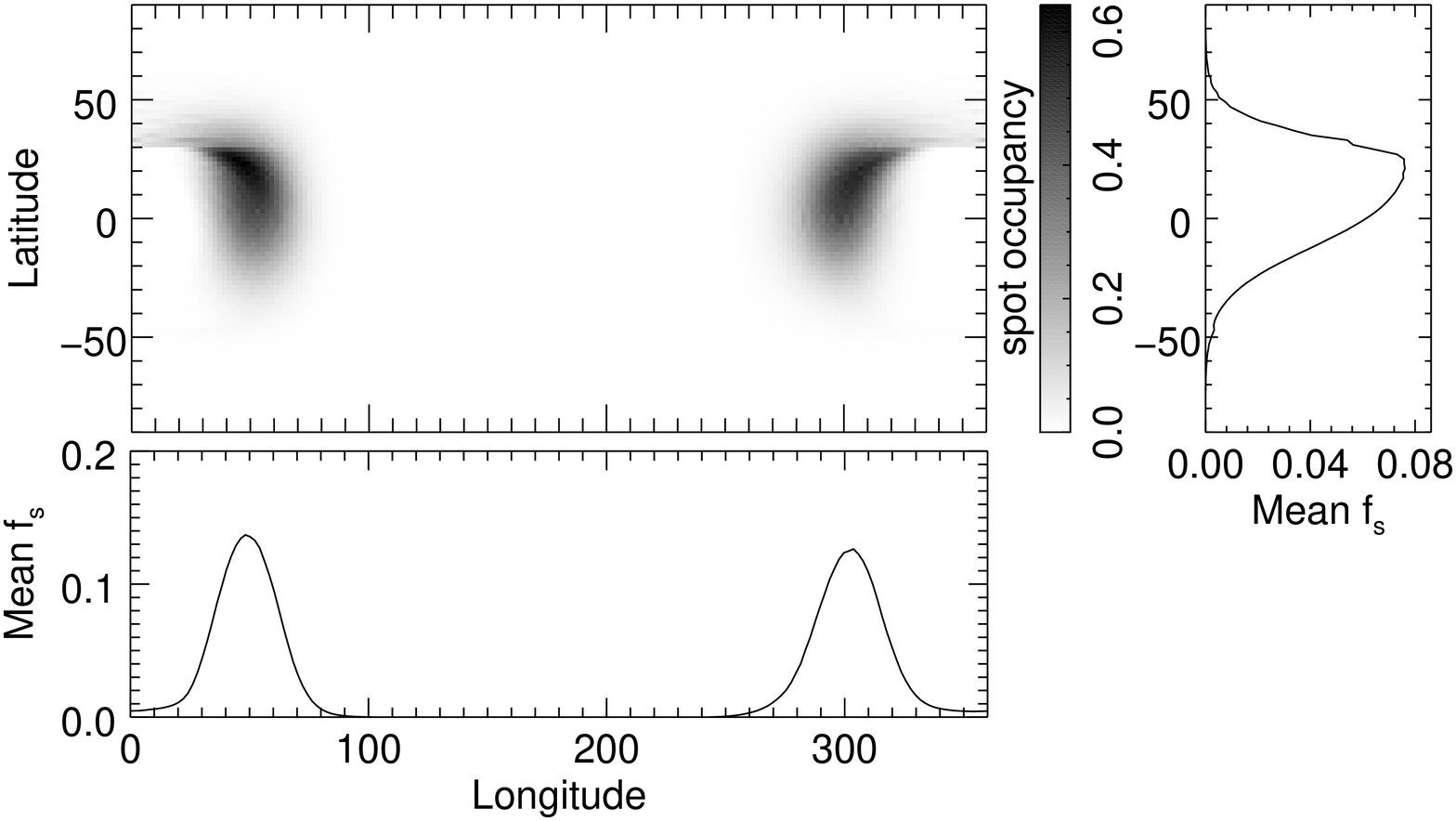}
\end{tabular}
\begin{tabular}{c}
\includegraphics[width=18.0pc]{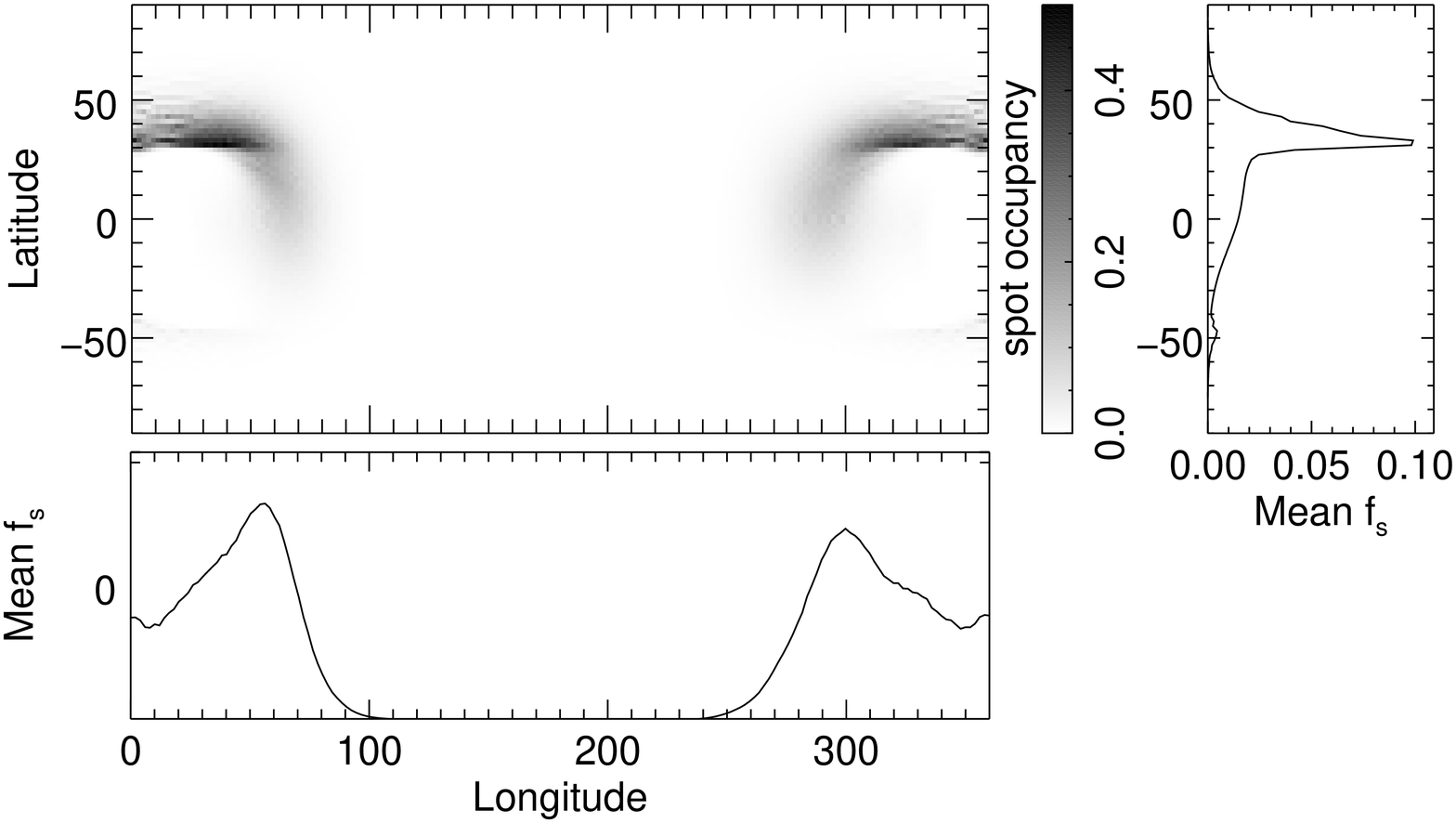}
\end{tabular}
\begin{tabular}{c}
\includegraphics[width=18.0pc]{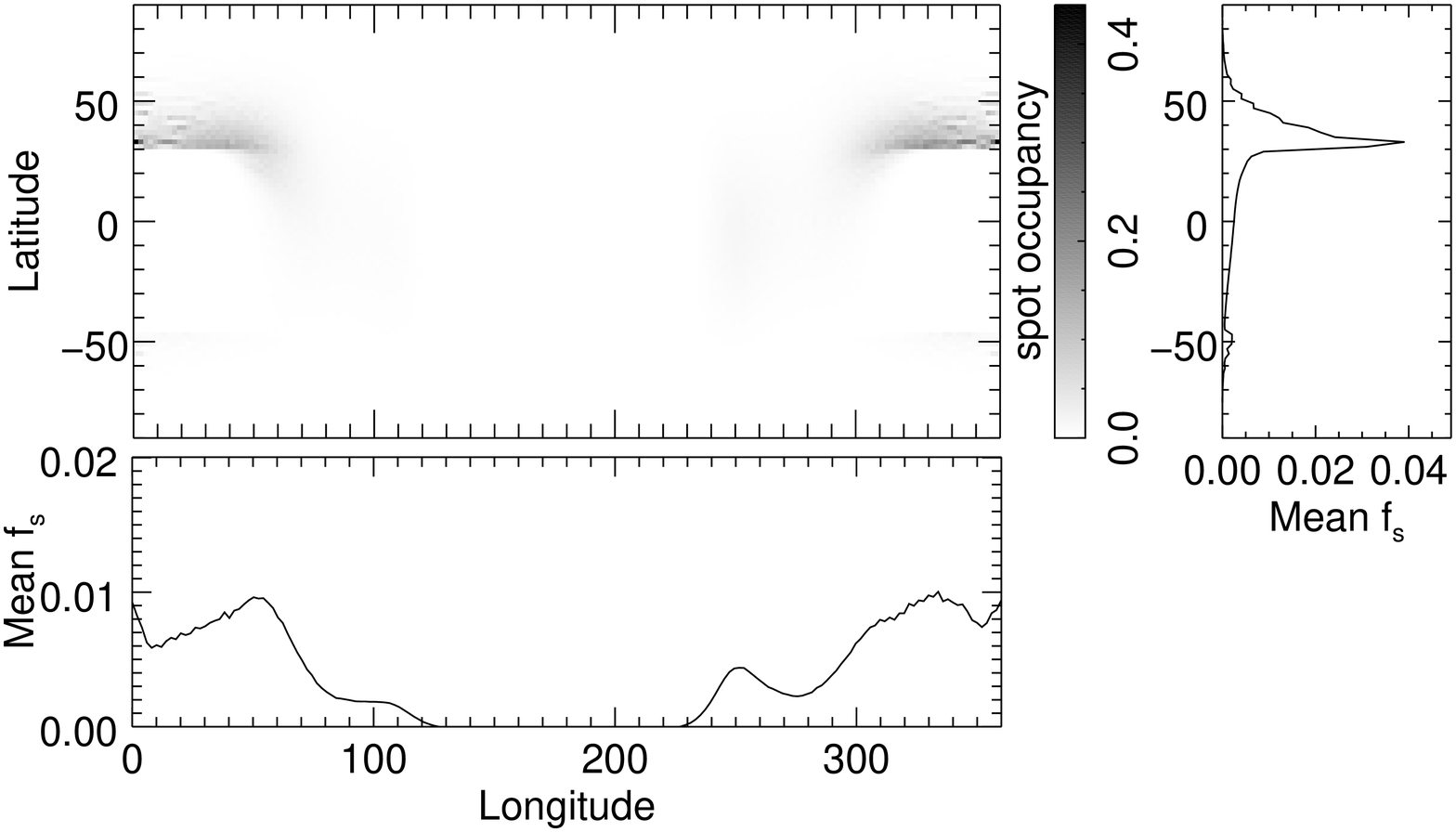}
\end{tabular}
\caption{Test results concerning the reconstruction of spots with different latitudes from top to bottom: 0$^{\circ}$, 20$^{\circ}$,  60$^{\circ}$ and 80$^{\circ}$.}
\label{fig:testlat}
\end{figure}

In order to investigate this pattern, we performed numerical tests. We generated four sets of synthetic spotted light curves using the same system parameters and the data sampling of SV Cam light curves. We put two spots with the same size (r = 20$^{\circ}$) and contrast. We fixed the longitudes to represent our map of SV Cam, as $\lambda_{Sp1}$ = 50$^{\circ}$ and $\lambda_{Sp2}$ =300$^{\circ}$. We considered four cases for the spot latitude: $\varphi_{case1}$ = 0$^{\circ}$, $\varphi_{case2}$ = 20$^{\circ}$, $\varphi_{case3}$ = 60$^{\circ}$ and $\varphi_{case4}$ = 80$^{\circ}$. The reconstructed images are shown in Figure~\ref{fig:testlat}. With increasing input latitude, the reconstructed spots become more asymmetric about the equator, as a consequence of vertical elongation for orbits with high inclination. For $\varphi\geqslant 20$, a sharp peak starts to appear at around latitude 30$^{\circ}$, where the limb of the secondary component traces the boundary with the occulted part of the primary. 
Our R-band reconstruction in Fig.~\ref{fig:phomap} resembles a 
linear combination of the cases 2 and 3, indicating that both high- and low-latitude spots are present on the primary star. 
This experiment demonstrates that the structure as well as the latitudinal distribution of reconstructed spots may potentially be used to constrain the latitudes of spots on the eclipsed components of binary systems \citep{Cameron1997}.

\subsection{Chromospheric activity}

We were able to reconstruct the surface map of the primary component only, since the contribution of the secondary component to the total flux is considerably low ($\sim 11\%$, depending on the orbital phase and wavelength). However, the high activity at the chromospheres of both components of SV Cam is also known from the literature (e.g., \citealt{Montes1995}, \citealt{Pojmanski1998}, \citealt{Ozeren2001}, \citealt{Kjurkchieva2002}). Therefore, thanks to the wide wavelength coverage and good quality of our spectra, we were also able to investigate the behaviour of the chromospheric activity in SV~Cam by means of the spectral subtraction technique. This method, which is simply based on the comparison of active (target) and non-active (standard) star spectra, was first suggested and applied by \citet{Barden1985}. In the optical wavelength range, it can be successfully applied to chromospheric activity indicators such as the CaII H\&K, H$\alpha$, and H$\beta$ lines, obtaining chromospheric line emission features as residuals. 

\subsubsection{Spectral synthesis}

\begin{figure}
\centering
	\includegraphics[trim=1 10 1 50,clip,width=\columnwidth]{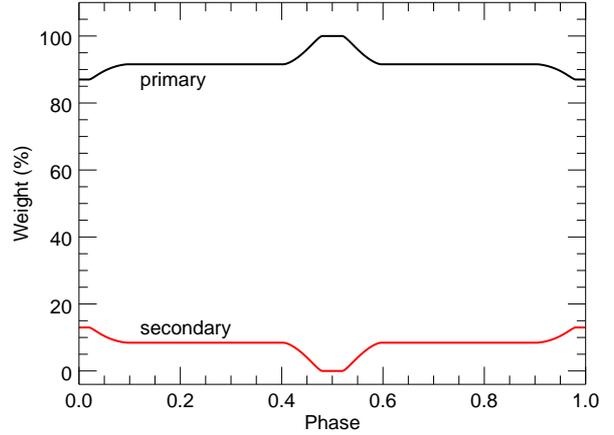}
 	\caption{The weights of each component of SV Cam in the H$\alpha$ region as a function of the orbital phase. The black solid line represents the weight of the primary component, while red solid line shows that of the secondary component.}
    \label{fig:fluxcont}
\end{figure}

Since SV Cam is an eclipsing binary, the spectra of the system include flux contribution from both components whose intensity and RV depend on the orbital phase. The variations in RV and flux contribution must be taken into account when constructing a non-active template spectrum. To this purpose, we improved a simple light curve model based on the equations given by Dan Bruton\footnote{http://www.physics.sfasu.edu/astro/ebstar/ebstar.html} to calculate the visible area of each component seen by the observer throughout the orbital phase. This simple model assumes spherical components revolving around circular orbits and neglects limb / gravity darkening effects. For the computations, we used the physical parameters of SV~Cam ($M_{1,2}$, $R_{1,2}$, $R_{1,2}$ \textit{a}, \textit{i}), which we derived from the simultaneous light and RV analysis of the system. The wavelength-dependent contribution of each component to the total continuum is calculated using the parameter 
\begin{equation}
\alpha :=\left (\frac{R_{1}}{R_{2}}\right)^{2}\left ( \frac{B_{1}(\lambda )}{B_{2}(\lambda )} \right )\left ( \frac{A_{1}(\phi)}{A_{2}(\phi)} \right ),
\label{eq:lumrat}
\end{equation}
where $R_{1,2}$ are the radii, $B_{1,2}(\lambda)$ are the Planck functions, and $A_{1,2}(\phi)$ represent  the fractional projected area of each component as seen by the observer at phase $\phi$ (0 when fully eclipsed, 1 for outside eclipse). Consequently, the weights for the primary and secondary components ($W_{1,2}$) are given by
\begin{equation}
W_{1}=\frac{\alpha }{1+\alpha }\hspace{0.5cm};\hspace{0.5cm}	W_{2}=\frac{1 }{1+\alpha }
\label{eq:weights}
\end{equation}
which are represented in Figure \ref{fig:fluxcont} for the H$\alpha$ region.

We performed spectral subtraction to La Palma data only, because it has a better phase coverage, allowing a more precise RV solution. To represent the non-active contribution of SV~Cam, we used the spectra of 18~Sco (G2V) and 61~CygA (K5V) as low-activity templates for the primary and the secondary components, respectively. In fact, 61 Cyg A is known to be a moderately active star \citep{Duncan1991}. \citet{Boro2016} reported the presence of a possible magnetic cycle that is twice the length of the activity cycle and also observed polarity reversals. On the other hand, 61 Cyg A has been used as a template star several times in the literature for spectral subtraction technique \citep[see, e.g.,][]{Martinez2010, Alonso2015}. It is also considered as a low-activity star \citep[see Table 2 of][]{Martinez2010}, in comparison with other similar mid-K type dwarf stars. The \textit{vsini} value of 61 Cyg A is $\sim$ 1.1 km/s \citep{Marsden2014}. During our spectral synthesis of SV Cam, the spectra are broadened at the \textit{vsini} value of the secondary component of SV Cam, 69 km/s. Therefore, the effect of activity in the core of the H$\alpha$ and Ca II IRT lines become much lower and may be considered as negligible. Since we compare the variation of photospheric (DI) and chromospheric activity (spectral synthesis) of SV Cam, the effect of the activity on the H$\alpha$ and Ca II IRT lines of 61 Cyg A only leads to an offset on the equivalent width axis. This does not affect the essence of our results, as we do not intend to provide standardised equivalent widths of SV Cam.

\begin{figure}
\centering
	\includegraphics[trim=1 10 1 50,clip,width=\columnwidth]{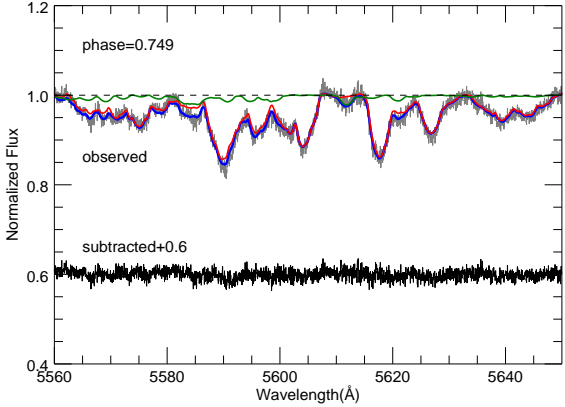}
 	\caption{The results of the consistency test for the spectral subtraction method for SV Cam. The gray solid line represents the normalized spectrum of SV Cam corresponding to phase 0.749. The red and green solid lines show the spectra of the primary and secondary components, respectively, while the blue solid line represents the total flux. The black solid line shows the residuals from the fit.
}
    \label{fig:contsub}
\end{figure}

The spectra of 18~Sco and 61~CygA were also obtained during La Palma observing run. The weights for the two components at the phases of SV~Cam observations were calculated using the relative flux contribution mentioned above (Eqs.~\ref{eq:lumrat}
and \ref{eq:weights}). The rotational equatorial velocities were calculated assuming rotation periods equal to the orbital one (synchronous rotation) for both components and the radii listed in Table~\ref{tab:table2}. We obtained $V_{eq}^{1}$ = 112 km/s and $V_{eq}^{2}$ = 69 km/s for the primary and secondary component, respectively. The spectral subtraction method also requires a careful normalization process. Since the standards used for the spectral subtraction were also obtained with the HERMES spectrograph and the data reduction was performed with the automatic pipeline, the normalization of each selected chunk including the activity indicator lines is easily performed, using second-order polynomial fits. To test the consistency of the spectral synthesis with the measured radial velocities, the calculated relative weights, and \textit{vsini} values of both components, we first modelled a spectral region which includes mostly photospheric lines (between 5560 $\mbox{\AA}$ and 5650 $\mbox{\AA}$). To this purpose, we used the spectrum of SV Cam taken at phase 0.749, in which the separation between the components is maximum. The resultant fit (total flux) as well as the residuals (subtracted spectrum) are shown in Figure \ref{fig:contsub}. The match between the observed and synthetic spectra is well enough to apply the spectral subtraction technique. 

\subsubsection{Variations in subtracted spectra}

\begin{figure*}
\begin{multicols}{2}
    \includegraphics[trim=1 10 1 40,clip,width=\columnwidth]{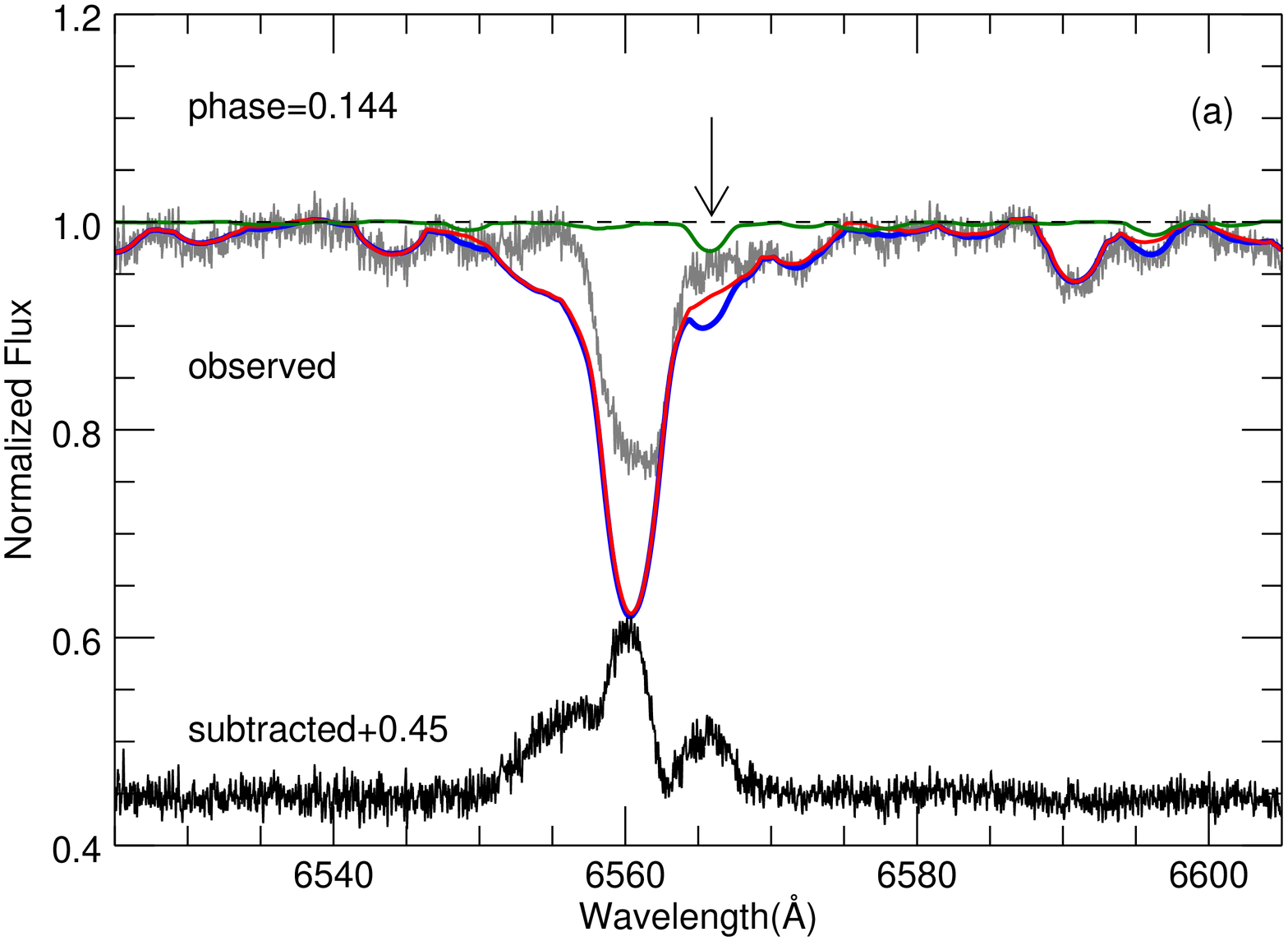}\par 
    \includegraphics[trim=1 10 1 40,clip,width=\columnwidth]{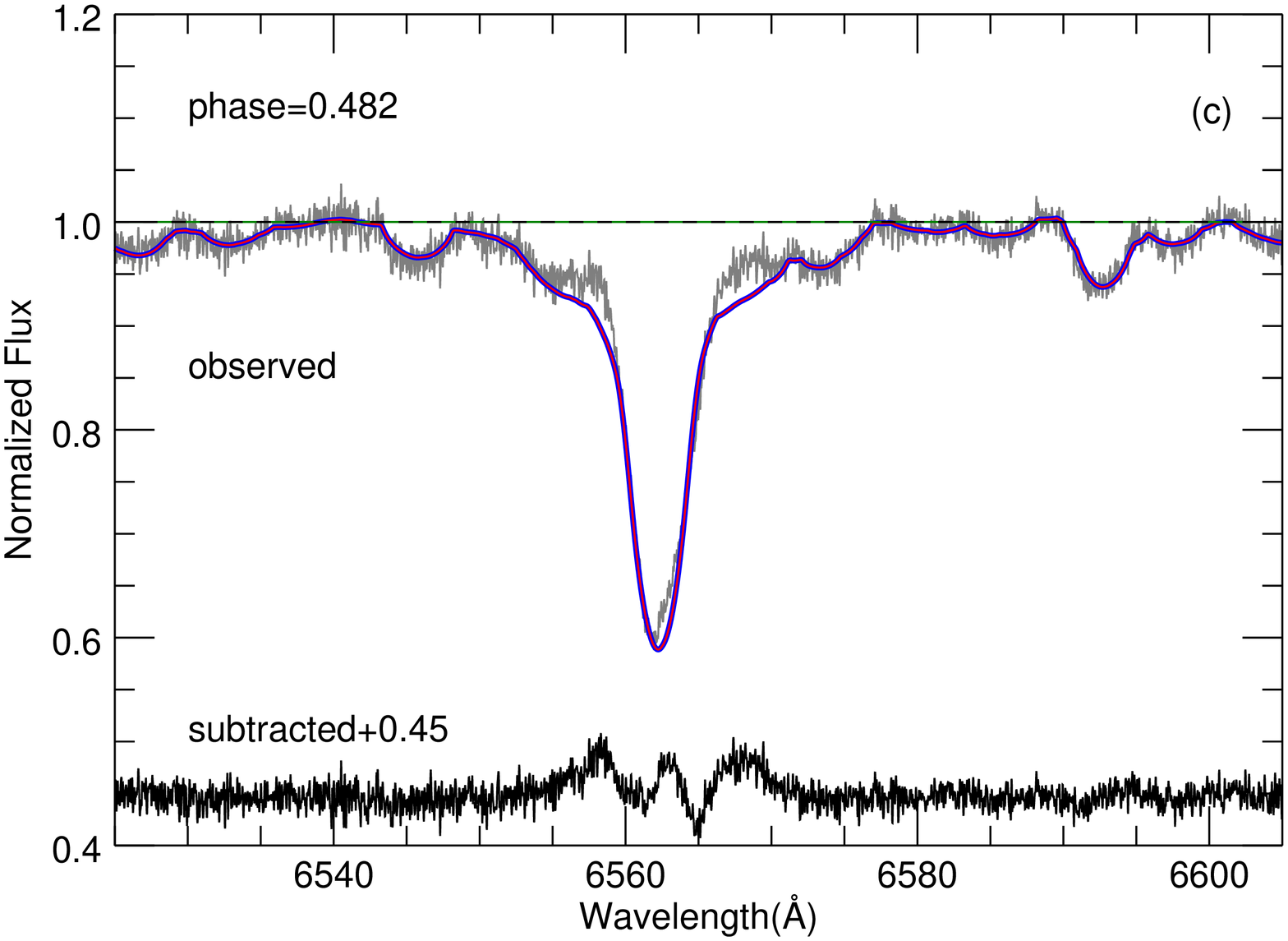}\par 
    \includegraphics[trim=1 10 1 40,clip,width=\columnwidth]{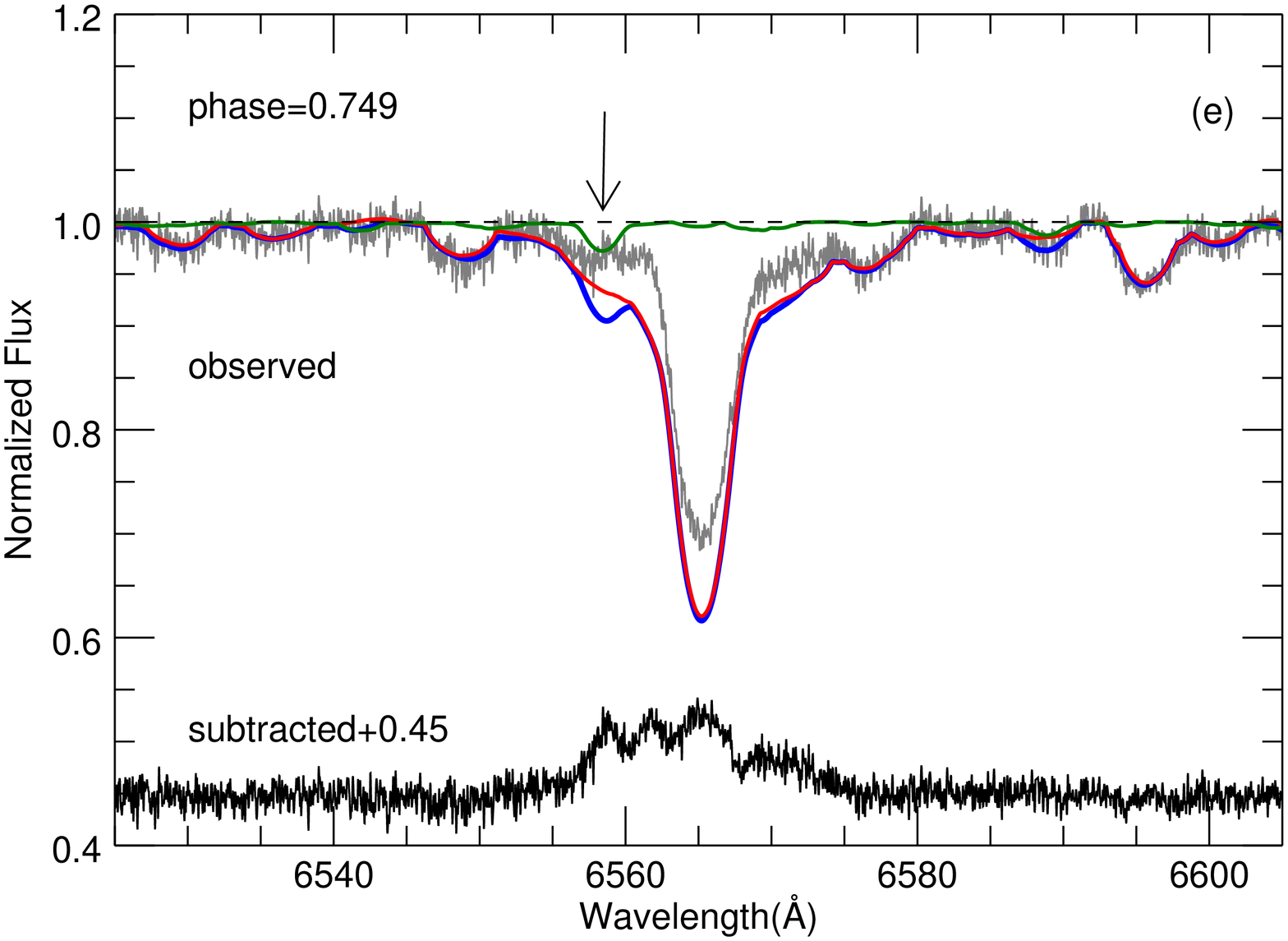}\par 
    \includegraphics[trim=1 10 1 40,clip,width=\columnwidth]{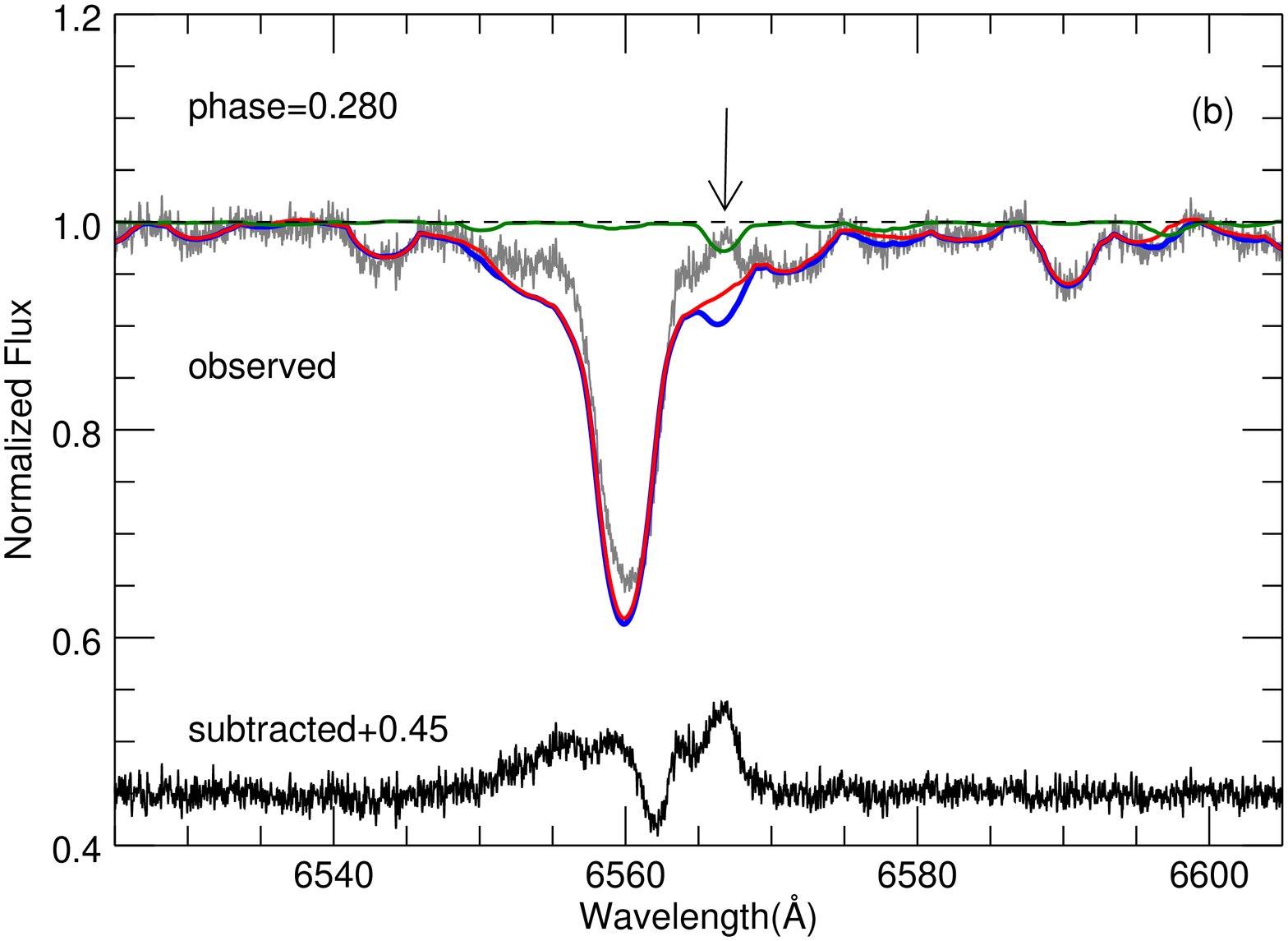}\par 
    \includegraphics[trim=1 10 1 40,clip,width=\columnwidth]{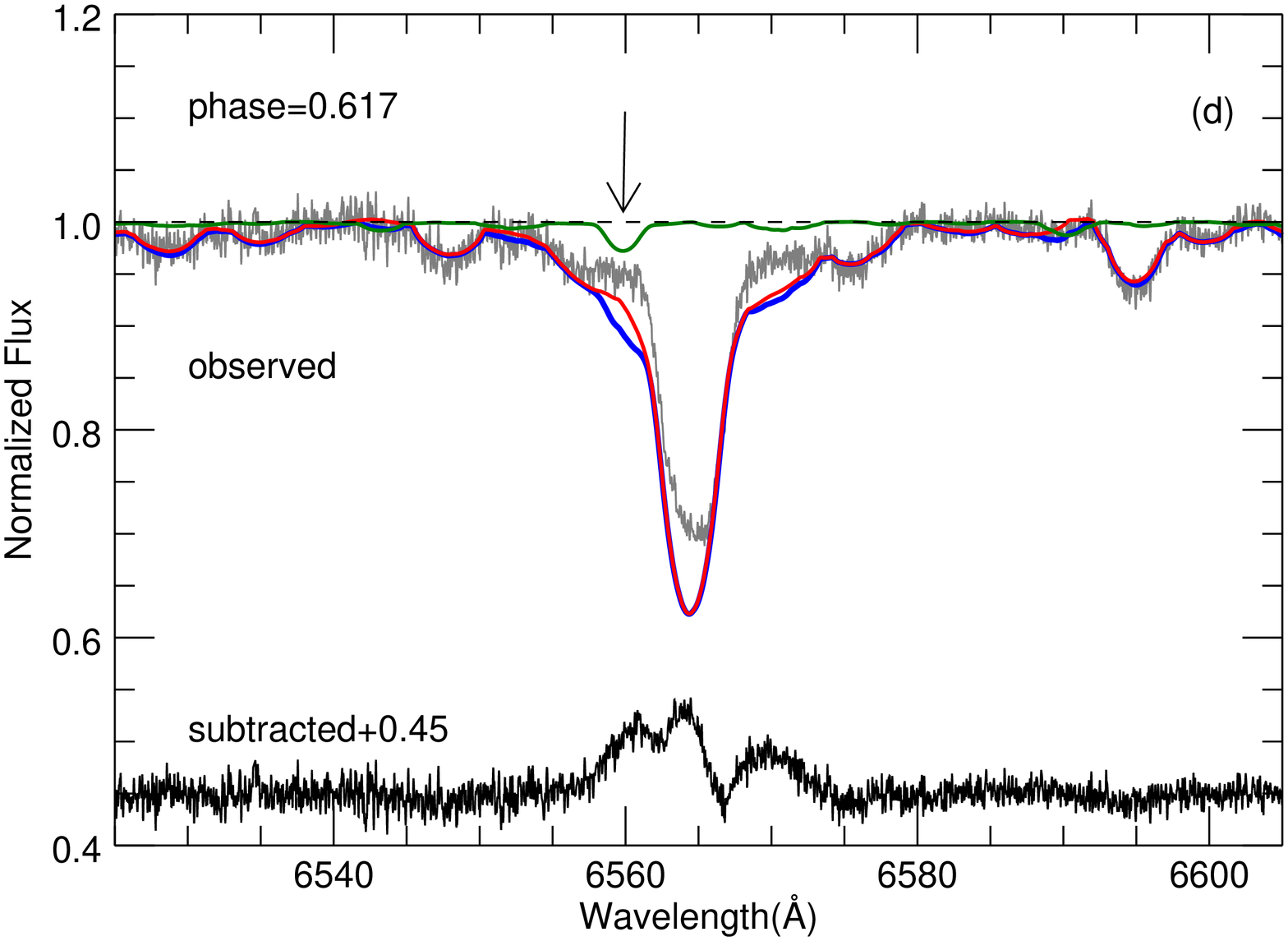}\par 
    \end{multicols}
\caption{The H$\alpha$ variation of SV Cam together with the synthetic spectra and the residuals. The arrows show the positions of H$\alpha$ lines of the secondary component. The color coding is as same as Fig. \ref{fig:contsub}.}
    \label{fig:halphas}
\end{figure*}

After showing the validity of synthetic spectra, we first investigate the H$\alpha$ line behaviour of the system with the help of the spectral subtraction technique, using the HERMES data from La Palma. The subtraction process is shown for a sample of phases in Fig.~\ref{fig:halphas}. The small amount of contribution of the secondary component to the total flux, along with the rapid rotation of the system prevent us to clearly disentangle the individual contribution of each component. In addition, the maximum wavelength separation between the components is $\sim$ 6 $\mbox{\AA}$, well below the line width of the primary. Therefore, when applying the spectral subtraction technique, we consider the total H$\alpha$ profile of the system, and integrated the area above or below the zero level in the subtracted profile, within $\pm 20$ $\mbox{\AA}$ about the line centre. 

\begin{figure}
\centering
	\includegraphics[trim=1 1 1 5,clip,width=\columnwidth]{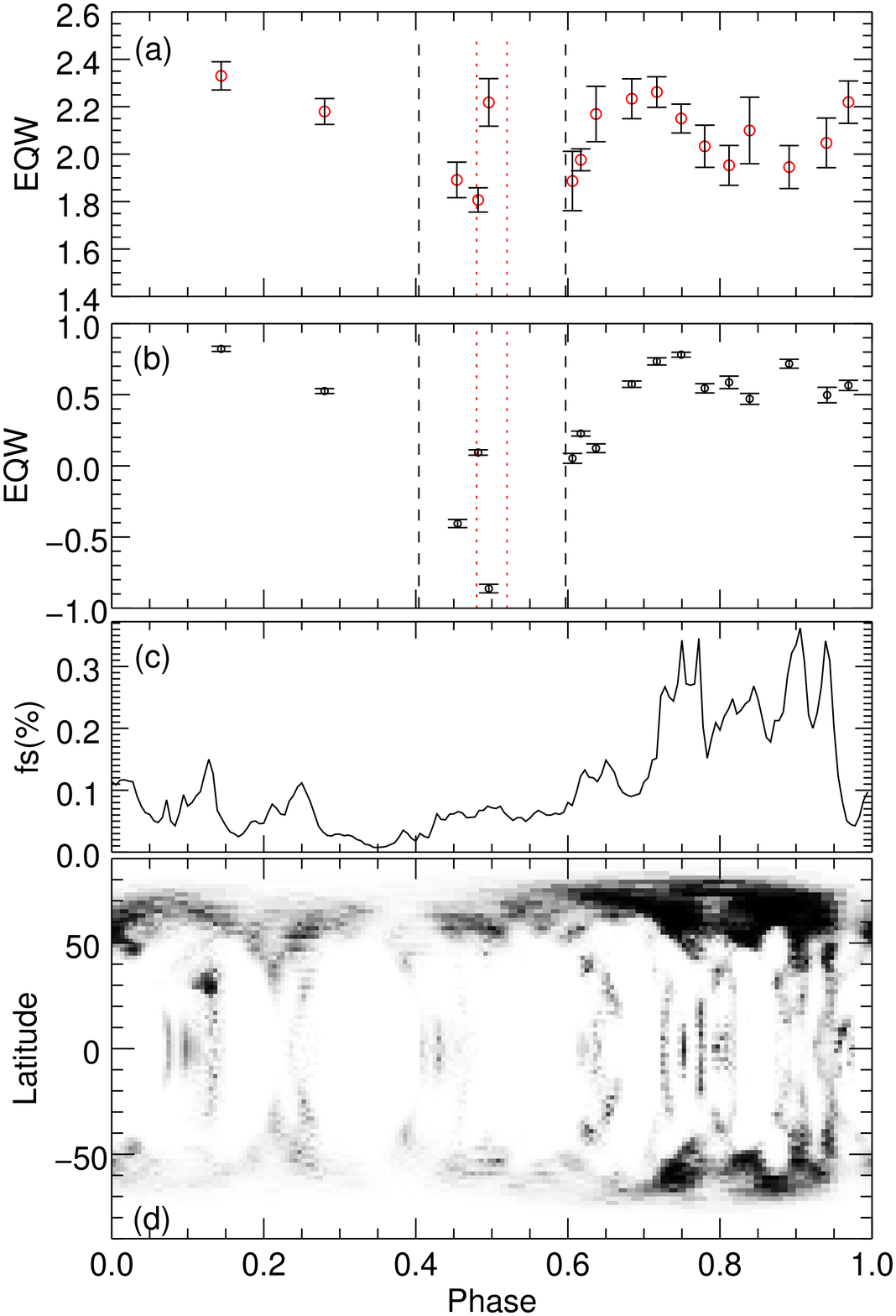}
 	\caption{The variation of the equivalent width of Ca II IRT excess emission (a) and H$\alpha$ excess emission/absorption (b), obtained using the spectral subtraction technique. The black dashed lines show the beginning and the end of the secondary minimum, while the red dotted lines represent the beginning and ending phases of the totality. The spot filling factor [$f_{s}$] (c) and the surface map of the primary component of SV Cam (d), obtained from the DI analysis.}
    \label{fig:halphacaemmap}
\end{figure}

The resulting phase variation of the excess H$\alpha$ equivalent width is shown in Figure \ref{fig:halphacaemmap}b, where positive and negative values represent excess emission and absorption, respectively. 
At the phases of large velocity separation (0.144 and 0.280), an emission feature is clearly visible in the residual spectra (Fig.~\ref{fig:halphas}), at the location of the secondary component. This unambiguously demonstrates the high activity level of the K4V-type secondary. 

Another interesting feature is the excess absorption seen around phase 0.5, when the primary star is in 
front of the secondary, which is clearly displayed in Fig.~\ref{fig:halphacaemmap}b. 
Extra-absorption in the H$\alpha$ line has been observed at phases preceding or following the eclipses in other RS~CVn stars and has been explained as the effect of prominences projected over the disc of the occulted star, similar to H$\alpha$ filaments on the Sun \citep[see, e.g.,][]{Hall1992, Frasca2000}. For SV~Cam, a similar excess absorption was reported by \citet{Ozeren2001} (see their Fig.~3) between phases 0.619 and 0.744.

Another good indicator of chromospheric activity is the Ca II infrared triplet (Ca II IRT) lines, which have several advantages over the Balmer lines, because Ca II IRT lines are less affected by the telluric absorptions and atmospheric extinction; they also have a well defined continuum \citep{Frasca2010}. Therefore, we have also investigated the variation of Ca II IRT lines for HERMES data, under the same assumptions considered for H$\alpha$, but within $\pm 12$~\AA~region. The resulting total variation of Ca II IRT excess emission, i.e. the sum of the excess equivalent widths of the three members of the triplet ($\lambda$8498, $\lambda$8542, and $\lambda$8662), is shown in Fig.~\ref{fig:halphacaemmap}a. The spot filling factor ($f_{s}$) variation along with the orbital phase and the corresponding DI map of the primary component are shown in Fig. ~\ref{fig:halphacaemmap}c and Fig. ~\ref{fig:halphacaemmap}d to track the spot distribution. Here, the variation of $f_{s}$ and excess H$\alpha$ is in accordance with each other especially between phases 0.5 and 1.0, while in the phase range 0.0--0.5 the $f_{s}$ values are lower but the excess H$\alpha$ emission is high. However, we point out that the small number of spectra in this phase range may have caused a poorer DI reconstruction. A comparison of Fig. ~\ref{fig:halphacaemmap}a and Fig. ~\ref{fig:halphacaemmap}b clearly shows that the variation of Ca II IRT lines closely follows the H$\alpha$ trend, except for two spectra showing excess H$\alpha$ absorption, obtained during the secondary minimum. 

At phases 0.45 and 0.50, the H$\alpha$ line shows significant excess absorption, while the IRT maintains its excess emission below the average level. This indicates that the plage contribution from the primary star's chromosphere is more effective in filling up the IRT lines than absorption, whereas the H$\alpha$ line is very sensitive to relatively cool filaments hanging in the chromosphere. To check whether cool filaments have a measurable effect, we measured the flux ratio of the IRT lines at 8542$\mbox{\AA}$ to 8498$\mbox{\AA}$ and found that the ratio fluctuates between 1.26 and 1.86 through all orbital phases, with a mean value of 1.46. These values are indicative of optically thick emission which is likely the result of a dominant contribution of chromospheric plage regions associated with star-spots. This is supported by the spot distribution near the secondary minimum (Fig.~\ref{fig:halphacaemmap}d), which is not very different than the more densely sampled phases. We conclude that even when there is significant excess absorption in H$\alpha$, the chromospheric IRT emission integrated over the stellar disc dominates the effect of any optically thin material near the stellar disc.

\section{Conclusions}

\begin{itemize}
\item
To obtain accurate orbital and physical parameters of the system, we modelled multi-band light curves with high precision and short cadence, simultaneously with radial velocity data, which required spot modelling during the analysis. The best-fit system parameters are consistent with our photometric mapping and Doppler imaging results using the DoTS code.
\end{itemize}

\begin{itemize}
\item
The light curve inversion performed for the R-band light curve resulted in two huge spots centered at longitudes $\sim$50$^{\circ}$ and 300$^{\circ}$ of the primary (F9V). This is consistent with the the simultaneous multi-band light curve and RV solution. The inverted map of the primary component indicates that the strongest activity in 2016.00 was concentrated around the phases where the primary is facing the secondary component.
\end{itemize}

\begin{itemize}
\item The primary component of SV Cam shows a high coverage of spots, which can be found at all longitudes.
Though the DI maps show that the spots on the primary component are concentrated at high latitudes excluding the poles, the numerical tests concerning the orbital inclination sensitivity of DI hints that they could actually be polar spots. The small amount of flux contribution of the secondary component prevents us from obtaining the DI maps of that companion.
\end{itemize}

\begin{itemize}
\item
The spectral synthesis and subtraction applied to this close binary system clearly reveals strong chromospheric activity of the secondary (K4V) component, though its filling fraction in the systemic radiative flux is very low. The phase variations of H$\alpha$ and Ca II IRT lines indicate that the primary (F9V) is covered with strong chromospheric plage regions throughout all the phases, in parallel with spot regions. The excess H$\alpha$ absorption detected near the secondary eclipse can be led by large cool prominence structures overcoming the background excess emission. 
\end{itemize}

\section*{Acknowledgements}

The authors are grateful to the anonymous referee for critical comments, which helped to improve the manuscript substantially. HV\c{S} acknowledges the support by The Scientific And Technological Research Council Of Turkey (T\"{U}B\.{I}TAK) through the project 1001 - 115F033. DM acknowledges support by the Spanish Ministry of Economy and Competitiveness (MINECO) 
from project AYA2016-79425-C3-1-P. AF acknowledges Istituto Nazionale di Astrofisica (INAF) for financial support. EI acknowledges support by the Young Scientist Award Programme BAGEP-2016 of the Science Academy, Turkey.




\bibliographystyle{mnras}
\bibliography{senavci_etal_rev2.bib} 

\bsp	
\label{lastpage}
\end{document}